\PassOptionsToPackage{unicode}{hyperref}
\PassOptionsToPackage{hyphens}{url}
\documentclass[
]{article}
\usepackage{amsmath,amssymb}
\usepackage{lmodern}
\usepackage{iftex}
\ifPDFTeX
  \usepackage[T1]{fontenc}
  \usepackage[utf8]{inputenc}
  \usepackage{textcomp} 
\else 
  \usepackage{unicode-math}
  \defaultfontfeatures{Scale=MatchLowercase}
  \defaultfontfeatures[\rmfamily]{Ligatures=TeX,Scale=1}
\fi
\IfFileExists{upquote.sty}{\usepackage{upquote}}{}
\IfFileExists{microtype.sty}{
  \usepackage[]{microtype}
  \UseMicrotypeSet[protrusion]{basicmath} 
}{}
\makeatletter
\@ifundefined{KOMAClassName}{
  \IfFileExists{parskip.sty}{%
    \usepackage{parskip}
  }{
    \setlength{\parindent}{0pt}
    \setlength{\parskip}{6pt plus 2pt minus 1pt}}
}{
  \KOMAoptions{parskip=half}}
\makeatother
\usepackage{xcolor}
\IfFileExists{xurl.sty}{\usepackage{xurl}}{} 
\IfFileExists{bookmark.sty}{\usepackage{bookmark}}{\usepackage{hyperref}}
\hypersetup{
  hidelinks,
  pdfcreator={LaTeX via pandoc}}
\usepackage{graphicx}
\makeatletter
\def\maxwidth{\ifdim\Gin@nat@width>\linewidth\linewidth\else\Gin@nat@width\fi}
\def\maxheight{\ifdim\Gin@nat@height>\textheight\textheight\else\Gin@nat@height\fi}
\makeatother
\setkeys{Gin}{width=\maxwidth,height=\maxheight,keepaspectratio}
\makeatletter
\def\fps@figure{htbp}
\makeatother
\ifLuaTeX
  \usepackage{selnolig}  
\fi

\author{}
\date{}

\begin{document}

Combinatorial Discovery of Irradiation Damage Tolerant Nano-structured
W-based alloys

Haechan Jo \textsuperscript{a}, Sanghun Park \textsuperscript{a}, Daegun
You \textsuperscript{a}, Sooran Kim \textsuperscript{b}, Dongwoo Lee
\textsuperscript{a, *}

\textsuperscript{a} School of Mechanical Engineering, Sungkyunkwan
University (SKKU), Suwon 16419, Republic of Korea

\textsuperscript{b}Department of Physics Education, Kyungpook National
University (KNU), Daegu 41944, Republic of Korea

*Corresponding Authors Email: dongwoolee@skku.edu

\textbf{Abstract}

One of the challenges in fusion reactors is the discovery of plasma
facing materials capable of withstanding extreme conditions, such as
radiation damage and high heat flux. Development of fusion materials can
be a daunting task since vast combinations of microstructures and
compositions need to be explored, each of which requires trial-and-error
based irradiation experiments and materials characterizations. Here, we
utilize combinatorial experiments that allow rapid and systematic
characterizations of composition-microstructure dependent irradiation
damage behaviors of nanostructured tungsten alloys. The combinatorial
materials library of W-Re-Ta alloys was synthesized, followed by the
high-throughput experiments for probing irradiation damages to the
mechanical, thermal, and structural properties of the alloys. This
highly efficient technique allows rapid identification of composition
ranges with excellent damage tolerance. We find that the distribution of
implanted He clusters can be significantly altered by the addition of Ta
and Re, which play a critical role in determining property changes upon
irradiation.


\emph{\textbf{\hfill\break
}}

1. Introduction

One of the technological challenges of commercialization of fusion
reactors from a materials perspective is the development of suitable
plasma facing materials (PFMs) that can withstand severe environments,
such as irradiation damages and huge heat flux {[}1{]}. It is therefore
required to investigate composition-microstructure dependent irradiation
damages of PFM candidates. One of the most promising PFMs in a fusion
reactor is tungsten because of its high melting point (3680 K), high
thermal conductivity (173 W·m\textsuperscript{-1}·K\textsuperscript{-1})
{[}2{]}, low sputtering yield {[}3{]}, low tritium retention {[}4{]},
high heat load resistance {[}5{]}, and high-temperature strength
{[}6{]}. However, PFMs are adversely affected by the irradiation of the
energetic neutrons and ions, which creates microstructural defects
{[}7{]}. Typical microstructural features of neutron-irradiated W are
the formation of dislocation loops, voids, and irradiation-induced
precipitates {[}8{]}. For hydrogen irradiation, supersaturation of
hydrogen leads to blistering in the near-surface region. This phenomenon
occurs even at a low irradiation energy range {[}9{]}. Moreover,
implanted helium (He) ions are readily trapped by irradiation-induced
defects, such as vacancies, dislocation loops, and voids resulting in
the formation of He-bubbles {[}10{]}, blisters {[}11{]}, and even fuzzes
{[}12{]}. These irradiation-induced defects and morphology changes cause
severe degradation of W, leading to increased tritium retention
{[}13{]}, a high erosion rate {[}14{]}, and decreased heat transfer
capacity.

Previous studies have shown that irradiation damages can be mitigated by
alloying with small amounts of additives to W. For example, W-Ta alloys
not only exhibit higher ultimate tensile strength (UTS) and prevent
crack propagation {[}15{]}, but also suppress the formation of He
bubbles {[}16{]}, and fuzzes {[}17{]}. W-Re alloys are known to improve
ductility {[}18{]}, and effectively suppress defects formed by neutron
irradiation {[}19{]}. Meanwhile, recent studies have revealed that
nanostructured tungsten and its alloys and composites, which contain a
high grain boundary (GB) density, are irradiation damage tolerant since
GBs serve as effective trapping sites for irradiation-induced defects
{[}20-22{]}. For instance, He-ion irradiated nanochannel W films
effectively release He atoms to GBs, resulting in a lower areal density
of He bubbles {[}23-25{]}. Similarly, S Si et al. reported that
W-graphene nanostructured multilayers exhibit excellent radiation
tolerance as the inserted graphene layers create new interfaces that act
as effective sinks for He-induced defects {[}26{]}. O EI-Atwani et al.
reported nanocrystalline W-based refractory high entropy alloys (HEAs)
with outstanding radiation resistance under Kr-ion irradiation {[}27{]}.
The W-based HEAs did not contain irradiation-induced dislocation loops
and had negligible irradiation hardening. Processing techniques have
been developed to fabricate the ultra-fine grained W at large scales
{[}28{]} and deposition techniques were suggested to produce
nanostructured coatings for PFMs {[}29, 30{]}

Although nanostructured W-alloys seem promising as PFMs, limited
combinations of composition-microstructure have been explored.
Irradiation damages of vast combinations of composition and
microstructure need to be investigated in order to deepen our
understanding on the irradiation damage tolerant materials and to
discover novel alloys with the ideal set of properties. Trial-and-error
based conventional irradiation experiments followed by materials
characterizations, however, are inefficient to investigate the huge
design space. In this work, combinatorial synthesis of nanostructured
W-Ta, W-Re, and W-Re-Ta thin films was carried out to produce specimens
with composition spreads. We then employed combinatorial and
high-throughput experiments (HTEs) to investigate irradiation damages of
the materials library {[}31-34{]}. The irradiation damage behaviors with
respect to composition and microstructure were characterized by
mechanical, thermal, and structural properties of the combinatorial
samples. We demonstrate from this highly efficient technique that
composition-microstructure combinations with the least irradiation
damages can readily be discovered.

2. Experimental

\textbf{2.1 Combinatorial film preparation}

The nanocrystalline W-based alloys were magnetron sputter deposited on a
475 $\mu$m thick Si (100) wafer capped with a 200 nm \
thick LPCVD \(\text{Si}_{3}N_{4}\) layer. The substrate was cut into a
\(20\ \text{mm}\  \times 20\ \text{mm}\) in-plane size for the W-Re-Ta
systems and into a \(20\ mm\  \times 10\ mm\) size for the W-Re and W-Ta
systems. The sputter depositions were carried out without a substrate
rotation to synthesize thin films with in-plane composition gradients
{[}32{]}. The sputter guns and the substrate are tetragonally arranged
such that different positions of the substrate have different relative
fluxes of the vapors from each target. The pure W film was also
deposited without a substrate rotation to induce a similar nanocolumnar
structure to the alloy systems. For the sputter deposition process, the
DC gun powers were set to 70 W, 20 W, and 30 W for W, Re, and Ta
targets, respectively, to prepare for the alloy specimens. A gun power
of 100 W was used for the pure W film. The depositions were carried out
at a substrate temperature of $500 \,^{\circ}\mathrm{C}$, and the Ar pressure and the base
pressure of the vacuum chamber were 5 \(\times\) 10\textsuperscript{-3}
Torr and 2 \(\times\) 10\textsuperscript{-7} Torr, respectively. Before
all the deposition processes, a 10 nm thick Ti adhesion layer was
deposited (DC power: 100 W, substrate rotation speed: 10 rpm, substrate
temperature: $500 \,^{\circ}\mathrm{C}$, Ar and base pressures: 5 \(\times\)
10\textsuperscript{-3} and 2 \(\times\) 10\textsuperscript{-7} Torr).

\textbf{2.2 Film characterizations}

The in-plane chemical composition gradient of the combinatorial films
was determined using an Energy-dispersive X-ray Spectroscopy (EDS) in a
Field Emission Scanning Electron Microscope (FESEM, JSM-7600F, JEOL).
The spot size used was \(1\ \mu m \times 1\ \mu m\) and the accelerating
voltage was set to 15 keV. The film thickness values of the
combinatorial samples were measured to be \(1.0 \pm 0.2\ \ \mu m\) using
a profilometer (DXT-A, Bruker) and confirmed with the cross-sectional
TEM and SEM micrographs.

X-ray diffraction (XRD) measurements were performed in the
Bragg-Brentano geometry (D8 advanced eco, Bruker). The film lattice
parameter and FWHM were determined using the (110) peak. The lattice
parameter was calculated through the d-spacing value using Bragg's law.
For the crystallinity analysis, the FWHM value of the gaussian-fitted
peak was measured. The cross-sectional transmission electron microscope
(JEM-2100F, JEOL) images were taken for both un-irradiated and
irradiated samples (at 200 keV). The 50 \textasciitilde{} 70 nm thick
TEM samples were prepared by using a focused ion beam milling (3D FEG,
Quanta) {[}35-37{]} with 30 kV Ga-ions (7 nA \textasciitilde{} 50pA).
Before the milling process, a protective layer of Pt was deposited to
prevent ion damages.

The electronic thermal conductivity (the electronic contribution of the
thermal conductivity) was determined by the Wiedemann-Franz (W-F) law
using electrical resistivity, which was measured by a custom-built
4-point measurement. A 4-point probe was configured in a line array with
1 mm of a spacing. The film electrical resistance (\emph{R}) was
determined using a current input of 1 mA through a digital multimeter
(3058e, RIGOL). Electrical resistance (\emph{R}) was converted to
resistivity (\(\rho\)) using film thickness (t) and a correction factor
(F = 4.23 \textasciitilde{} 4.36) {[}38, 39{]} :

\(\ \ \ \ \ \ \ \ \rho = \ FRt\). (1)

The electronic thermal conductivity (\(\kappa_{e}\)) was calculated
through the W-F law {[}40{]}:

\(\frac{\kappa_{e}}{\sigma} = LT,\) (2)

where \(\sigma\) is electrical conductivity (\(1/\rho\)), \emph{L} is
the Lorenz number (\(2.44 \times 10^{- 8}W\Omega K^{- 2}\)), and
\emph{T} is the absolute temperature. The thermal conductivity
determined by this method does not consider the thermal transport by
phonons, which constitutes approximately 10 \% of the total thermal
conductivity for metals and alloys {[}40-42{]}.

The hardness maps were determined by nanoindentation (KLA iNano) in the
continuous stiffness measurement (CSM) mode and with a diamond Berkovich
tip. Before running the experiments, a standard tip calibration was
performed using a fused silica sample. The target load and strain rate
used were 50 mN and 0.2 \(s^{- 1}\), respectively. The nanoindentation
hardness values versus depth were determined using the Oliver \& Pharr
method {[}43{]}. We took the values at a 10\% depth of the film
thickness to avoid the substrate effect {[}44, 45{]}. A total of 16
positions per composition were analyzed (\(4 \times 4\) array, 25 $\mu$m intervals). The combinatorial synthesis method used in
this work resulted in surface roughness of 20 nm, approximately. The
measured hardness would be slightly underestimated by the sample
roughness (Fig. S1), but would be slightly overestimated by the
compressive residual stresses of the thin film.

\textbf{2.3 First-principles calculations}

First-principles calculations based on density function theory (DFT)
were performed through the Vienna Ab initio simulation package (VASP).
The generalized gradient approximation of Perdew-Burke-Ernzerhof has
been implemented. Host atom valence states were treated using -W, Re,
and Ta with PAW-PBE W\_pv, Re\_pv, Ta\_pv, and the interstitial impurity
atom -- He with PAW-PBE He. The energy cutoff of 520 eV and a
\(5 \times 5 \times 5\) mesh of Monkhorst-pack points for the Brillouin
zone integration were employed for the calculation.

3. Results

\includegraphics[width=4.5in]{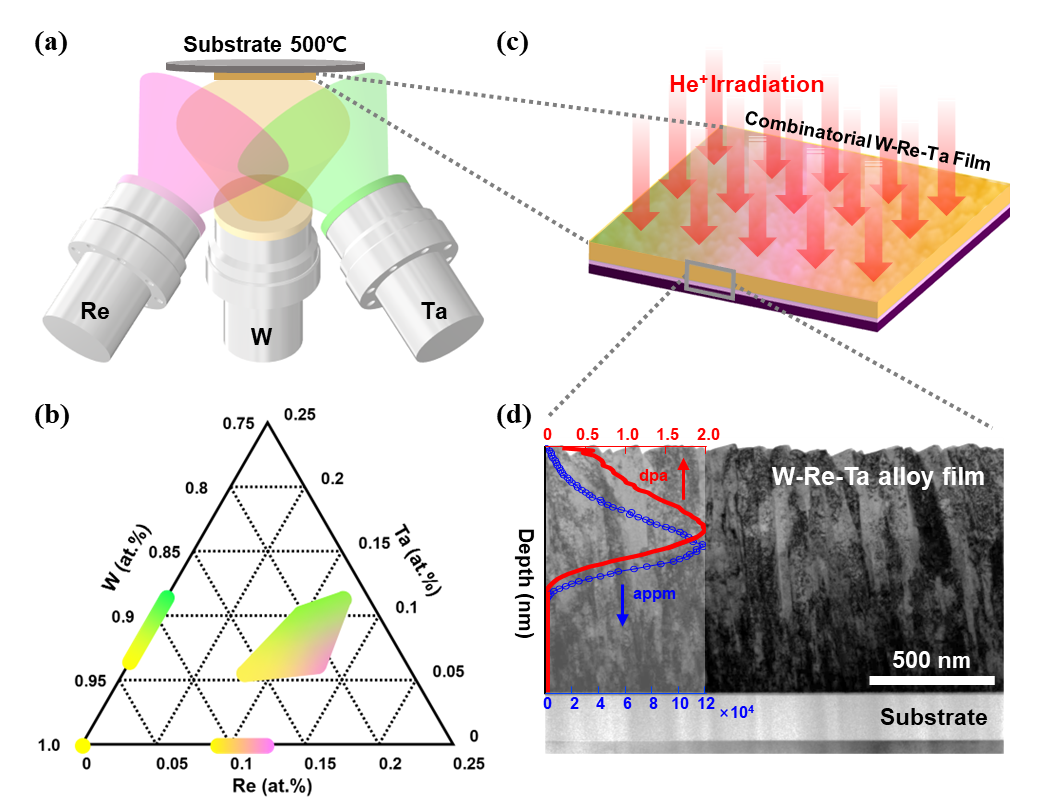}

Fig. 1. (a) A schematic diagram of the combinatorial synthesis
processing using magnetron sputtering for the nanostructured binary and
ternary alloys with W, Re, and Ta. (b) Composition ranges of the
nanostructured alloys investigated in this work. (c) An illustration of
the He-ion irradiation experiment conducted on the combinatorial
specimen. (d) A cross-sectional TEM image of a W-Re-Ta alloy film.
Displacement per atom (dpa) and He\textsuperscript{+} ion concentration
versus depth calculated by SRIM 2013 are shown.

Pure W and binary (W-Re and W-Ta), ternary (W-Re-Ta) nanostructured
alloys with composition spreads were prepared by a magnetron co-sputter
deposition process with a substrate temperature of 500 ℃, as
schematically illustrated in Fig. 1(a). During the deposition, the
substrate was not rotated, producing 1 $\mu$m thick thin film specimens with in-plane chemical composition gradients.
The composition ranges investigated in this work is shown in Fig. 1(b),
which were determined by energy dispersive X-ray spectroscopy (EDS):
pure W, W\textsubscript{87-91}Re\textsubscript{9-13},
W\textsubscript{88-94}Ta\textsubscript{6-12}, and
W\textsubscript{75-87}Re\textsubscript{8-13}Ta\textsubscript{5-12}
ranges were synthesized, which correspond to 50 distinct compositions
with a 1 at. \% spacing. Two specimens for each system under the same
deposition condition were prepared, one of which was subjected to the
mechanical, electrical, and microstructure characterizations in the
un-irradiated state. The other specimen was used for the He-ion
irradiation experiment, followed by the identical materials
characterizations.

All the combinatorial specimens prepared in this work can be located
within a 40 mm \(\times\) 40 mm area together, where a uniform
irradiation condition can be assured by the He-ion irradiation source
(KOMAC, Korea Multi-Purpose Accelerator Complex). He-ions with an energy
of 200 keV and the fluence and flux of 2\(\times\)10\textsuperscript{17}
ions\(\bullet\)cm\textsuperscript{-2} and
1.1\(\times\)10\textsuperscript{13}
ions\(\bullet\)cm\textsuperscript{-2}s\textsuperscript{-1}, respectively
were irradiated to the specimens in the growth direction of the films
(see Fig. 1(c)). The irradiation experiment was performed in a vacuum
chamber (\textasciitilde1.0\(\times\)10\textsuperscript{-5} Torr) at
room temperature. The damage profile calculated by Stopping and Range of
Ions in Matter 2013 (SRIM 2013) using the experimental irradiation
conditions employed and with the displacement threshold energy of 90 eV
{[}46, 47{]} is shown in Fig. 1(d), together with a cross-sectional
transmission electron microscopy (TEM) image of a W-Ta-Re nanostructured
alloy. The damage profile shows a maximum of 2 dpa (displacement per
atom) and 12 at. \% He concentration at a depth of \textasciitilde{} 380
\(\text{nm}\). Low homologous temperatures (the ratio of the deposition
temperature to the melting point of the specimen) of the thin films
resulted in a densely packed columnar nanostructure, which is consistent
with the T-zone microstructure of the structure-zone model {[}48{]}. The
nanocolumnar structure is expected to exhibit high damage tolerance as
high-density grain boundaries can act as sinks for irradiation
byproducts {[}49-52{]}.

\includegraphics[width=4.8in]{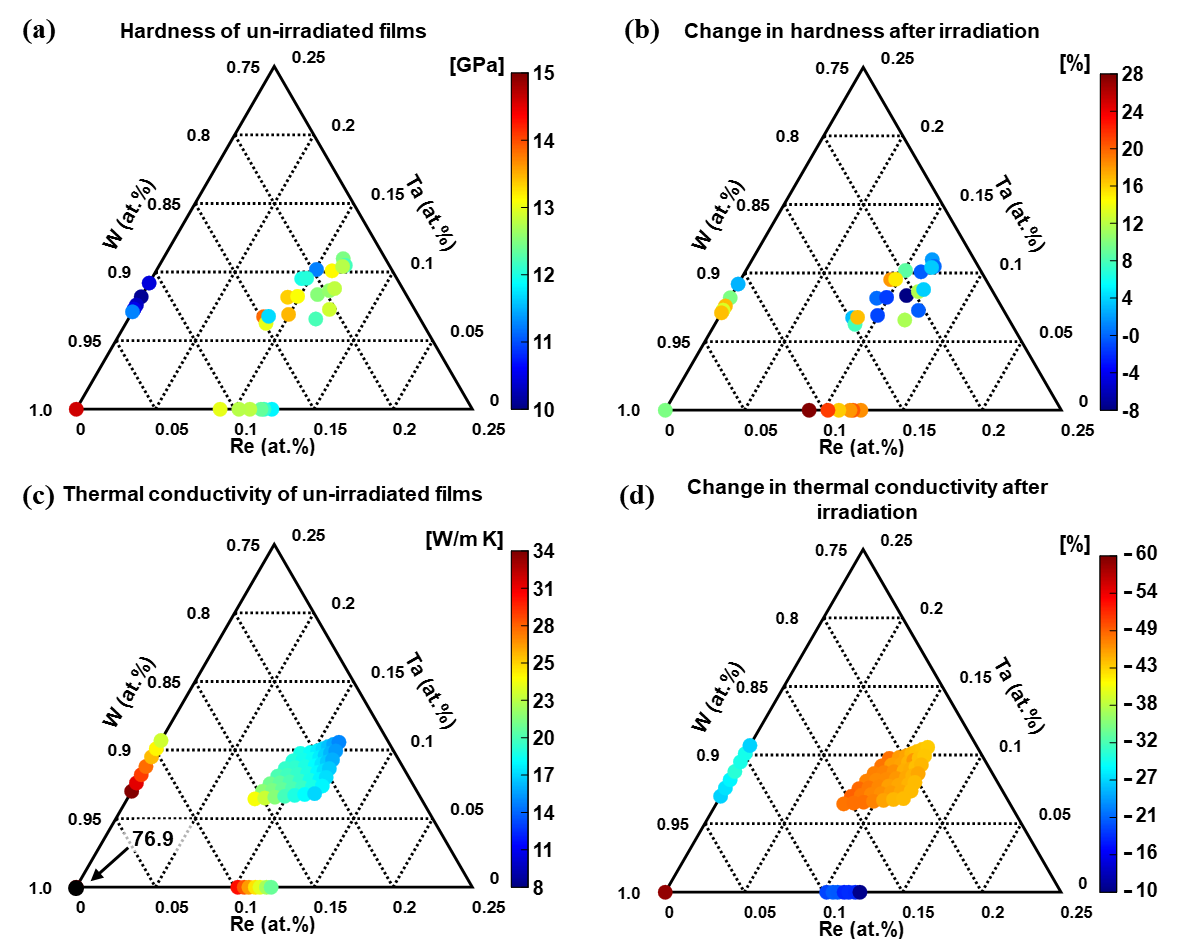}

\textbf{Fig. 2.} (a) Hardness map of the un-irradiated combinatorial
films. (b) The change in hardness of the specimens after He-ion
irradiation. (c) Thermal conductivity map of the un-irradiated films.
(d) The change in thermal conductivity after irradiation.

The irradiation-induced hardening of PFMs is one of the major
degradation mechanisms, which is closely related to the reliability of
fusion reactors. We used nanoindentation to investigate
composition-dependent degradation in the mechanical property of the
combinatorial nanostructured alloys. As shown in Fig. 2(a), the
nanoindentation hardness values of the un-irradiated alloy thin films of
the W-Ta (10.1 \textasciitilde{} 10.5 GPa) and W-Re (11.8
\textasciitilde{} 13.0 GPa) are lower than the hardness of the pure W
(14.6 GPa). It has been reported that Re solutes in W enhance the
mobility of screw dislocations, resulting in solid solution softening
{[}53-55{]}. On the other hand, Ta solutes in W generally increase
hardness due to grain refinement and relative density improvement during
bulk processing {[}56, 57{]}. As will be discussed later, the W-Ta
system in this work has a relatively large grain size than the other
systems, resulting in lower hardness values in our experiment {[}58{]}.

Irradiation-induced hardening have been characterized by comparing the
hardness maps acquired before and after He-ion irradiation (Fig. S2). As
seen in Fig. 2(b), the change in hardness is more severe in the W-Re
system (18 \textasciitilde{} 28 \%), than in the pure W (9.8 \%),
W-Re-Ta (-8 \textasciitilde{} 18 \%), and W-Ta (2.8 \textasciitilde{} 17
\%) films. It is noticeable from the result that the composition ranges
with small hardness changes are readily identifiable. For instance, the
W\textsubscript{91}Ta\textsubscript{9} and
W\textsubscript{76-83}Re\textsubscript{10-13}Ta\textsubscript{7-11}
composition regions have irradiation hardening rates below 4.9 \%, which
is half of the value of pure W (9.8 \%).

A similar irradiation damage map can be acquired for the thermal
property of the combinatorial thin film library. Fig. 2(c) shows the
electronic thermal conductivity (\(\kappa_{e}\)) map of the
un-irradiated combinatorial films, which was determined using the
electrical resistivity map and through the Wiedemann-Franz (W-F) law
(Fig. S3). The pure W film has the largest value of \(\kappa_{e}\) (76.9
W/m·K) and increment of solute concentration decreases \(\kappa_{e}\) as
the solutes scatter the electron flow {[}56, 59{]}. Ta lowers
\(\kappa_{e}\) more effectively than Re:
W\textsubscript{90}Ta\textsubscript{10} has 23.9 W/m\(\cdot\)K, while
W\textsubscript{90}Re\textsubscript{10} has 29.5 W/m\(\cdot\)K for
\(\kappa_{e}\). The result is somewhat unexpected because the grain size
of the W-Ta system is larger than the W-Re alloys and bulk Ta (14.6
\(\mu\Omega \bullet cm\) at $20\,^{\circ}\mathrm{C}$) has lower resistivity than bulk
Re (22.2 \(\mu\Omega \bullet cm\) at $20\,^{\circ}\mathrm{C}$) {[}60{]}. We attribute
lower thermal conductivity in the W-Ta system than in the W-Re system to
the larger degrees of the lattice distortion in W-Ta {[}61{]}, as will
be confirmed later. Since the W-Re-Ta films have higher overall solute
concentrations than the binary films, they have lower electronic thermal
conductivity.

After He-ion irradiation, \(\kappa_{e}\) was reduced for all the systems
(see Fig. 2(d)) due to enhanced scattering of electron flow due to the
irradiation-induced defects in the GBs and GIs (grain interiors).
Reduction in \(\kappa_{e}\) is larger in the pure W film (- 60 \%) than
in the alloys with Re and/or Ta solutes, indicating that the solutes
make the alloys more irradiation damage tolerant. Interestingly, the
W-Re system that has the most severe mechanical irradiation damage
showed the least degradation in \(\kappa_{e}\) among the investigated
systems. As will be discussed later, this result is due to different
distributions of He-irradiation induced defects in the W-Ta and W-Re
systems. Composition ranges with small irradiation damages in
\(\kappa_{e}\) can effectively be identified in Fig. 2(d): changes in
\(\kappa_{e}\) of W\textsubscript{89-93}Ta\textsubscript{7-11} and
W\textsubscript{87-90}Re\textsubscript{10-13} are less than half of the
thermal degradation in pure W (-60\%).

\includegraphics[width=4.8in]{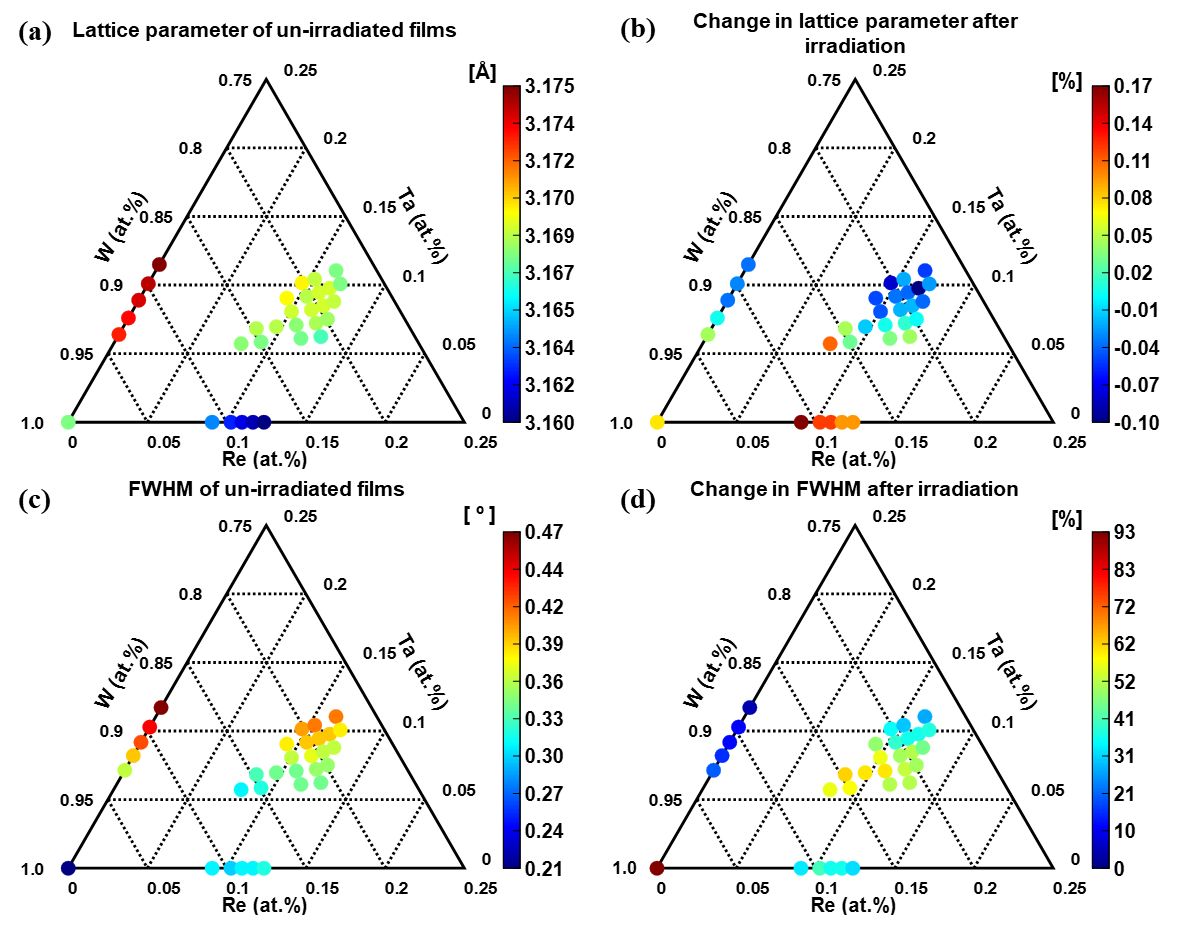}

\textbf{Fig. 3.} XRD results of the combinatorial nanostructured thin
films. The maps of (a) the lattice parameter of the un-irradiated films
and (b) the change in the lattice parameter due to He irradiation. (c)
The map of the FWHM from the (110) XRD peaks of the un-irradiated films
and (d) the change in the FWHM due to He irradiation.

Microstructural analyses were carried out through a wafer scanning X-ray
diffraction (XRD) technique for the un-irradiated and irradiated
combinatorial thin film alloys (Fig. 3). We find that all the
compositions of the W-alloys studied in this work formed a BCC phase
(space group: \(\text{Im}\overline{3}m\)) with a (110) preferred
orientation (Fig. S4). The lattice parameters of the un-irradiated
combinatorial films are presented in Fig. 3(a). In the binary systems,
increase in solute concentration in the W-Ta system leads to larger
lattice parameters, while a reverse trend is found in the W-Re system.
This trend is attributed to the relative atomic sizes of the solutes as
compared with W: the radius of Ta (1.47 Å) is largest, followed by W
(1.41 Å) and Re (1.38 Å) {[}62{]}. The combined effects of Ta
(increasing lattice parameter) and Re (decreasing lattice parameter) can
be seen for the W-Ta-Re ternary system.

The changes in the lattice parameters of the combinatorial thin films
after He-ion irradiation are presented in Fig. 3(b). The lattice
parameter map of the irradiated films can be found in Fig. S5(a). The
composition ranges with low irradiation damages in lattice parameter can
be readily identified using Fig. 3(b): changes in the lattice parameter
of W\textsubscript{89-92}Ta\textsubscript{8-12} and
W\textsubscript{77-84}Re\textsubscript{9-12}Ta\textsubscript{7-11} are
less than half of the lattice swelling in pure W. He-ion implantation
resulted in lattice swelling of the pure W (0.07 \%) and the W-Re films
(0.10 \textasciitilde{} 0.17 \%). Smaller lattice swelling or even
shrinkage is observed in the W-Ta and W-Re-Ta systems after irradiation
(- 0.10 \% \textasciitilde{} 0.11\%). There are two competing mechanisms
that are responsible for the different degrees of lattice swelling.
First, lattice shrinkage can occur by the irradiation-induced relaxation
of the non-equilibrium crystal structure that was formed during the
physical vapor deposition process {[}63-65{]}. Second, the insertion of
He atoms in the grains but outside of the He bubbles can make a lattice
expansion, whose rate would depend upon the solutes. RY Zheng et al.
{[}66{]} pointed out that most of the implanted He atoms in W do not
form He bubbles but may form very small He-vacancy complexes. When He
trapping sites are fully occupied, additional He atoms can also form
self-clusters {[}67-69{]}. In addition, formation of self-interstitial W
combined with interstitial He can occur after irradiation {[}70, 71{]}.
All these structural changes may attribute the lattice expansion. As a
simplified system to investigate the effects of solutes on the lattice
expansion, we consider W, W\textsubscript{89}Ta\textsubscript{11},
W\textsubscript{89}Re\textsubscript{11} supercells with different
concentrations (1.8 to 6.9 \%) of He interstitials at the tetrahedral
sites for DFT (density functional theory) simulations (\emph{Supporting
Information section 2}). We find that the
W\textsubscript{89}Ta\textsubscript{11} system experiences the smallest
lattice expansion after He insertion, followed by the W and
W\textsubscript{89}Re\textsubscript{11} (Fig. S10(b)). This trend is
consistent with the combinatorial experimental results in Fig. 3(b).

The periodicity of the atomic arrangement of the combinatorial specimens
has been characterized by determining the FWHM (full width at half
maximum) of the (110) peaks of the XRD patterns. Before He-ion
irradiation, the W-Re and W-Ta systems have FWHM ranges of 0.30 $\sim$ $0.32\,^{\circ}$ and 0.36 $\sim$ $0.47\,^{\circ}$, respectively,
while the value for the W film is $0.21\,^{\circ}$ (Fig. 3(c)). Therefore, the alloy systems show larger FWHM than pure W, indicating that the
insertion of Ta and Re solutes resulted in the reduction of the lattice
periodicity (lattice distortion) {[}72{]}. Because Ta solute has a
larger atomic radius than W (4.26 \%), a larger degree of lattice
distortion has occurred in the W-Ta system {[}73{]}. On the other hand,
alloying of W with Re resulted in less change in lattice periodicity
since Re atom is smaller than W, and there exists a smaller difference
in the atomic radii (2.13 \%) between the W and Re atoms. This trend can
also be confirmed in FHWM distribution in the un-irradiated W-Re-Ta
system (Fig. 3(c)): FWHM is strongly correlated with Ta concentration
but not with Re.

After irradiation, the FWHM in all systems increases probably due to the
formation of some defects such as vacancy, self-interstitial atoms
(SIAs), and He-clusters (Fig. S5(b)). The changes in the FWHM
\textbf{(}Fig. 3(d)\textbf{)} of the binary and ternary alloys (4.3
\textasciitilde{} 62.5 \%) are smaller than in pure W (93.1 \%),
suggesting that both the Ta and Re substitutional atoms are helpful for
suppressing changes in the periodicity of atomic arrangements. He
irradiation on the W-Ta system involves much smaller changes in FWHM
(4.3 \textasciitilde{} 20 \%) as compared to other systems. Also, as Ta
concentration in the W-Ta-Re alloys increases, the change in the FWHM
reduces. The experimental results on the structural periodicity can be
compared with the DFT simulation results shown in Fig. S10(c) and
(d)\textbf{.} That is, after the He insertion to the interstitial sites,
FWHM increases for all the supercells of W, W-Ta, and W-Re. Also, the
smallest change in FWHM by He is found in the W-Ta system as in the
experiment. Meanwhile, from the results of the combinatorial experiment,
the composition ranges with small irradiation damages in the FWHM are
readily identified in Fig. 3(d): changes in the FWHM of
W\textsubscript{88-91}Re\textsubscript{9-12},
W\textsubscript{88-93}Ta\textsubscript{6-12} and
W\textsubscript{77-82}Re\textsubscript{9-12}Ta\textsubscript{9-11} are
less than half of the pure W case. The intersection composition ranges
of low irradiation damages in the lattice parameter and the FWHM can be
identified as W\textsubscript{89-92}Ta\textsubscript{8-12} and
W\textsubscript{77-82}Re\textsubscript{9-12}Ta\textsubscript{9-11}.

\includegraphics[width=4.8in]{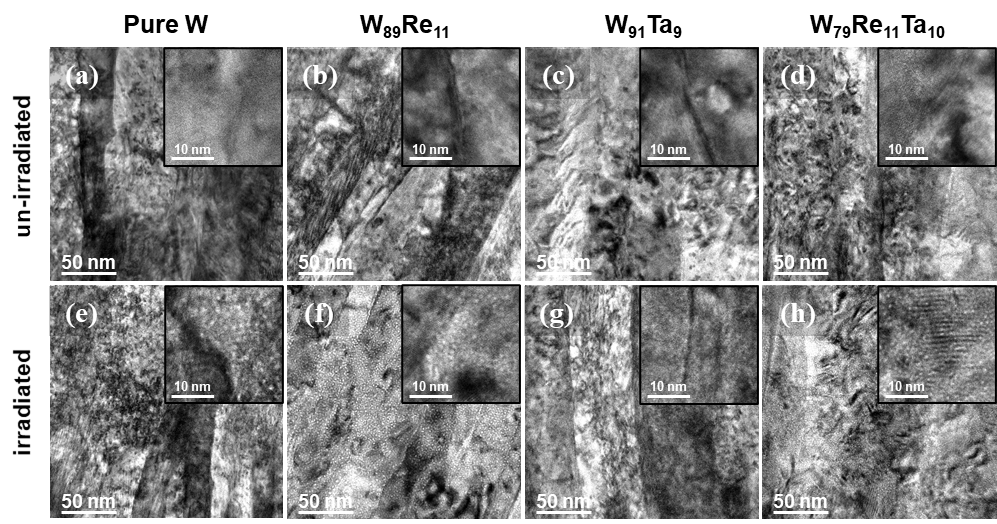}

\textbf{Fig. 4.} Cross-sectional bright-field TEM micrographs of (a, e)
Pure W, (b, f) W\textsubscript{89}Re\textsubscript{11,} (c, g)
W\textsubscript{91}Ta\textsubscript{9} and (d, h)
W\textsubscript{79}Re\textsubscript{11}Ta\textsubscript{10} thin film
alloys in (a-d) un-irradiated and (e-h) He-ion irradiated states. Insets
show corresponding images with a higher magnification.

Detailed microstructural features of the un-irradiated and irradiated
films of W, W\textsubscript{89}Re\textsubscript{11},
W\textsubscript{91}Ta\textsubscript{9}, and
W\textsubscript{79}Re\textsubscript{11}Ta\textsubscript{10} have been
investigated, as shown in Fig. 4 and Fig. S6-7. The cross-sectional TEM
images in Fig. 4 were obtained at a depth of \textasciitilde{} 380 nm of
the specimens, where the dpa and He concentrations are expected to be
their maxima (Fig. 1(d)). For all the specimens, dense nanocolumnar
structures were formed (Fig. S6). The in-plane grain sizes of the
specimens determined for un-irradiated W,
W\textsubscript{89}Re\textsubscript{11},
W\textsubscript{91}Ta\textsubscript{9}, and
W\textsubscript{79}Re\textsubscript{11}Ta\textsubscript{10} films are
\(53 \pm 25\ nm\), \(55 \pm 30\ nm\), \(89 \pm 35\ nm\),
\(57 \pm 25\ nm\), respectively. The largest grain size is found in
W\textsubscript{91}Ta\textsubscript{9}, which has the largest FWHM of
the XRD peak (Fig. 3(c)). Therefore, the large FWHM value in the
un-irradiated W-Ta system is not due to a small grain size but lattice
distortion as discussed earlier. Formation of large grains in the W-Ta
systems can be attributed to the misfit strain caused by the difference
in atomic radii of Ta and W, which leads to an enhanced grain growth
rate during the physical vapor deposition process {[}74{]}.

Fig. 4(e)-(h) displays the TEM micrographs of the He irradiated
specimens. The in-plane grain sizes of these specimens were determined
for pure W \((55 \pm 27\ nm)\), W-Re \((58 \pm 31\ nm)\), W-Ta
\((92 \pm 42\ nm)\) and W-Re-Ta \((63 \pm 27\ nm)\). The changes in the
grain sizes after irradiation are negligible. Therefore, the increase in
the FWHM after irradiation (Fig. 3(d)) is mainly due to the change in
the periodicity of the lattice, caused by the irradiation induced
defects. He bubbles are observed in the TEM images of the Pure W,
W\textsubscript{89}Re\textsubscript{11,} and
W\textsubscript{79}Re\textsubscript{11}Ta\textsubscript{10} films. Many
of large He bubbles are found in
W\textsubscript{89}Re\textsubscript{11}, while He bubbles might not
reach an observable size (\textasciitilde0.3 nm {[}75{]}) by TEM in
W\textsubscript{91}Ta\textsubscript{9}.

4. Discussion

The irradiation damage maps shown in Fig. 2(b), (d) and in Fig. 3(b),
(d) indicate that the addition of Ta in nanostructured W reduces
irradiation damages in microstructures and suppresses degradation in the
thermal and mechanical properties. The addition of Re also retards the
changes in electronic thermal conductivity and in the periodicity of the
atomic arrangement after He irradiation, but irradiated W-Re films
resulted in rather high irradiation hardening and lattice expansion
rates. Here, the effects of Ta and Re solutes in the nanostructured W on
the distributions of implanted He, and their roles in the change in the
physical properties are discussed.

\includegraphics[width=4.8in]{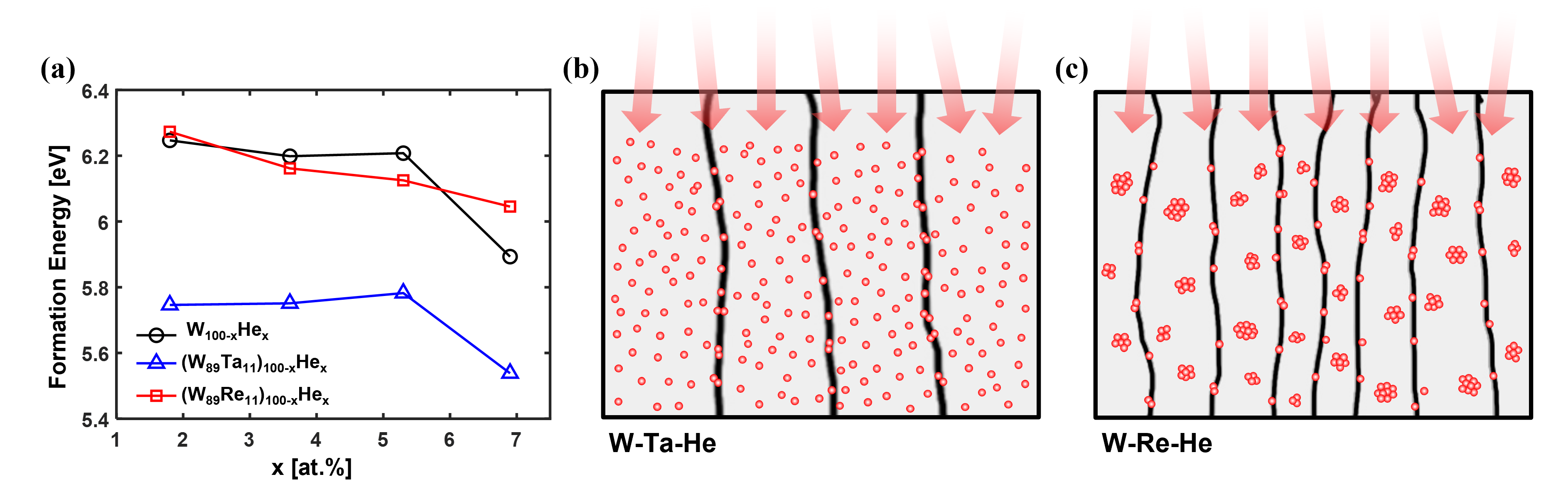}

\textbf{Fig. 5.} (a) Formation energies of the interstitial He atoms at
the tetrahedral sites in W, W\textsubscript{89}Ta\textsubscript{11}, and
W\textsubscript{89}Re\textsubscript{11} supercells, calculated by DFT
simulations. Schematic diagrams representing helium bubble formation in
the (b) W-Ta and (c) W-Re microstructures.

To investigate the effect of the solutes on the stability of implanted
He, we calculated the formation energies of interstitial He atoms at
random tetrahedral sites in pure W and
W\textsubscript{89}Ta\textsubscript{11},
W\textsubscript{89}Re\textsubscript{11} solid-solution supercells by
using the DFT simulations (\emph{Supporting Information section 2).} As
shown in Fig. 5(a), the interstitial He atoms have the lowest formation
energy and are thus most stable in the interstitial sites in the W-Ta
system. Also, it has been reported that the activation energy of He
diffusion is higher in W-Ta solid-solutions than in W or W-Re systems
{[}76, 77{]}. Based on these results, it is reasonable to conclude that
implanted He atoms are rather uniformly distributed with smaller cluster
sizes in the W-Ta system (Fig. 5(b)), while migration of He atoms and
the formation of larger clusters of them is expected for the W-Re system
(Fig. 5(c)). In this sense, larger He bubbles are expected to form in
the W-Re system, as can be seen from the TEM results (Fig. 4). Besides
the effects of the stability and diffusivity of He interstitials, there
are other effects to be considered to explain a preferred He bubble
formation in the alloys containing Re. Recent studies have shown that
the irradiated particles in W-Re alloys enhance the diffusivity of Re
atoms, leading to the formation of Re-clusters {[}78-80{]}. The Re
clusters have high binding energy with He, which enhances the nucleation
and growth of He bubbles {[}76{]}. Besides, Ta solutes in W-Re-Ta alloys
can suppress the formation of the Re-clusters {[}81, 82{]}. TEM
micrographs of
W\textsubscript{79}Re\textsubscript{11}Ta\textsubscript{10} in Fig. 4(h)
show that the addition of Ta in W-Re can reduce the He bubble formation
significantly.

Different sizes and distributions of He clusters or bubbles due to the
Re and Ta solutes lead to different degrees of degradation of the
mechanical property. Large He-bubbles in W-Re results in severe
irradiation hardening, because the defects interfere dislocation slips
effectively {[}83{]}. Lower hardening rate, or even softening occurs for
the alloys containing Ta, where a much smaller number of He bubbles were
formed. For these systems, both hardening and softening mechanisms are
expected to be activated: the smaller size and more uniform distribution
of He bubbles in W-Ta can suppress dislocation motion and can be
dislocation nucleation sources at the same time {[}84{]}. The electronic
thermal conductivity is also affected by the cluster sizes and the
distributions of implanted He in the alloys. Electron scattering occurs
more effectively in the specimens with uniformly distributed and
high-density small He bubbles (W-Ta, W-Re-Ta), as compared to the
samples with larger and lower density He bubbles (W-Re). This argument
can be supported by the molecular dynamics (MD) work on Si by T Wang et
al. which reported a more significant reduction in the thermal
conductivity by uniformly distributed vacancies than by the vacancies
clustering {[}85{]}.

5. Conclusion

In this study, combinatorial experiments for discovering He-irradiation
damage tolerant PFMs have been carried out. Combinations of the
combinatorial synthesis, a simultaneous He-ion irradiation experiment,
and the high-throughput characterizations were used to systematically
characterize the irradiation damages of a broad composition range of
nanostructured W-Ta-Re alloys. This set of techniques allows a facile
identification of the composition ranges with excellent damage tolerance
in mechanical (W\textsubscript{91}Ta\textsubscript{9} and
W\textsubscript{76-83}Re\textsubscript{10-13}Ta\textsubscript{7-11}) and
thermal (W\textsubscript{89-93}Ta\textsubscript{7-11} and
W\textsubscript{87-90}Re\textsubscript{10-13}) properties, as well as in
microstructure (W\textsubscript{89-92}Ta\textsubscript{8-12} and
W\textsubscript{77-82}Re\textsubscript{9-12}Ta\textsubscript{9-11}). It
has been revealed that the solutes of Re and Ta in nanostructured W
significantly affect the formation and distribution of He-bubbles, which
are closely related to the damage behaviors in microstructural and
physical properties. We believe that the suggested method can be used to
investigate a broader range of damage tolerant nanostructured alloys
with various irradiation conditions, facilitating the discovery of
advanced PFMs. Development of the bulk processing would be necessary for
the practical application of the nanostructured W-alloys as PFMs.

Acknowledgements

This research was supported by Basic Science Research Program through
the National Research Foundation of Korea (NRF) funded by the Ministry
of Science, ICT and Future Planning (NRF-2017R1E1A1A01078324,
NRF-2020M3D1A1016092).

\textbf{Data availability}

The raw and processed data required to reproduce these finding are
available from the corresponding author upon reasonable request.

\textbf{6. References}

{[}1{]} J. Knaster, A. Moeslang, T. Muroga, Materials research for
fusion, Nature Physics 12(5) (2016) 424-434.

{[}2{]} JW Davis, VR
Barabash, A Makhankov, L. Plochl, Assessment of tungsten for use in the
ITER plasma facing components, Journal of Nuclear Materials 258-263
(1998) 308-312.

{[}3{]} J.N. Brooks, L. El-Guebaly, A. Hassanein, T.
Sizyuk, Plasma-facing material alternatives to tungsten, Nuclear Fusion
55(4) (2015).

{[}4{]} R. Causey, K. Wilson, T. Venhaus, W.R. Wampler,
Tritium retention in tungsten exposed to intense fluxes of 100 eV
tritons, Journal of Nuclear Materials 266-269 (1999) 467-471.

{[}5{]} Z.
Zhou, J. Linke, G. Pintsuk, J. Du, S. Song, C. Ge, High heat load
properties of ultra fine grained tungsten, Journal of Nuclear Materials
386-388 (2009) 733-735.

{[}6{]} G.-M. Song, Y.-J. Wang, Y. Zhou, The
mechanical and thermophysical properties of ZrC/W composites at elevated
temperature, Materials Science and Engineering: A 334(1-2) (2002)
223-232.

{[}7{]} X. Hu, T. Koyanagi, M. Fukuda, N.A.P.K. Kumar, L.L.
Snead, B.D. Wirth, Y. Katoh, Irradiation hardening of pure tungsten
exposed to neutron irradiation, Journal of Nuclear Materials 480 (2016)
235-243.

{[}8{]} A. Hasegawa, M. Fukuda, K. Yabuuchi, S. Nogami, Neutron
irradiation effects on the microstructural development of tungsten and
tungsten alloys, Journal of Nuclear Materials 471 (2016) 175-183.

{[}9{]}
Y. Ueda, J. Coenen, G. De Temmerman, R. Doerner, J. Linke, V. Philipps,
E. Tsitrone, Research status and issues of tungsten plasma facing
materials for ITER and beyond, Fusion Engineering and Design 89(7-8)
(2014) 901-906.

{[}10{]} N. Yoshida, H. Iwakiri, K. Tokunaga, T. Baba,
Impact of low energy helium irradiation on plasma facing metals, Journal
of Nuclear Materials 337-339 (2005) 946-950.

{[}11{]} J. Yu, W. Han, Z.
Chen, K. Zhu, Blistering of tungsten films deposited by magnetron
sputtering after helium irradiation, Fusion Engineering and Design 129
(2018) 230-235.

{[}12{]} A. Al-Ajlony, J.K. Tripathi, A. Hassanein, Low
energy helium ion irradiation induced nanostructure formation on
tungsten surface, Journal of Nuclear Materials 488 (2017) 1-8.

{[}13{]}
Y. Nobuta, Y. Hatano, M. Matsuyama, S. Abe, Y. Yamauchi, T. Hino, Helium
irradiation effects on tritium retention and long-term tritium release
properties in polycrystalline tungsten, Journal of Nuclear Materials 463
(2015) 993-996.

{[}14{]} B.D. Wirth, K.D. Hammond, S.I. Krasheninnikov,
D. Maroudas, Challenges and opportunities of modeling plasma--surface
interactions in tungsten using high-performance computing, Journal of
Nuclear Materials 463 (2015) 30-38.

{[}15{]} M. Wirtz, J. Linke, G.
Pintsuk, L. Singheiser, I. Uytdenhouwen, Comparison of the thermal shock
performance of different tungsten grades and the influence of
microstructure on the damage behaviour, Physica Scripta T145
(2011).

{[}16{]} I. Ipatova, G. Greaves, S. Pacheco-Gutiérrez, S.C.
Middleburgh, M.J.D. Rushton, E. Jimenez-Melero, In-situ TEM
investigation of nano-scale helium bubble evolution in tantalum-doped
tungsten at 800°C, Journal of Nuclear Materials 550 (2021).

{[}17{]} S.
Gonderman, J.K. Tripathi, T.J. Novakowski, T. Sizyuk, A. Hassanein, The
effect of low energy helium ion irradiation on tungsten-tantalum (W-Ta)
alloys under fusion relevant conditions, Journal of Nuclear Materials
491 (2017) 199-205.

{[}18{]} Y. MUTOH., K. ICHIKAWA, K. NAGATA, M.
TAKEUCHI, Effect of rhenium addition on fracture toughness of tungsten
at elevated temperatures, Journal of Materials Science 30 (1995)
770-775.

{[}19{]} M. Fukuda, T. Tanno, S. Nogami, A. Hasegawa, Effects of
Re Content and Fabrication Process on Microstructural Changes and
Hardening in Neutron Irradiated Tungsten, Materials Transactions 53(12)
(2012) 2145-2150.

{[}20{]} Z. Chen, L.-L. Niu, Z. Wang, L. Tian, L.
Kecskes, K. Zhu, Q. Wei, A comparative study on the in situ helium
irradiation behavior of tungsten: Coarse grain vs. nanocrystalline
grain, Acta Materialia 147 (2018) 100-112.

{[}21{]} T. Lifeng, L. Pei, L.
Xuanze, M. Yutian, M. Xiangmin, Cracks and blisters formed in
nanocrystalline tungsten films by helium implantation, Fusion
Engineering and Design 172 (2021).

{[}22{]} I. Oh, D. Park, E. Cheong, H.
Jo, S. Park, D. You, T. Kim, Y. Park, K. Kim, G.-D. Sim, C. Shin, D.
Lee, Anisotropic He-ion irradiation damages in nanocolumnar W thin
films, Extreme Mechanics Letters 41 (2020).

{[}23{]} W. Qin, F. Ren, R.P.
Doerner, G. Wei, Y. Lv, S. Chang, M. Tang, H. Deng, C. Jiang, Y. Wang,
Nanochannel structures in W enhance radiation tolerance, Acta Materialia
153 (2018) 147-155.

{[}24{]} W. Qin, Y. Wang, M. Tang, F. Ren, Q. Fu, G.
Cai, L. Dong, L. Hu, G. Wei, C. Jiang, Microstructure and hardness
evolution of nanochannel W films irradiated by helium at high
temperature, Journal of Nuclear Materials 502 (2018) 132-140.

{[}25{]} W.
Qin, S. Jin, X. Cao, Y. Wang, P. Peres, S.-Y. Choi, C. Jiang, F. Ren,
Influence of nanochannel structure on helium-vacancy cluster evolution
and helium retention, Journal of Nuclear Materials 527 (2019).

{[}26{]}
S. Si, W. Li, X. Zhao, M. Han, Y. Yue, W. Wu, S. Guo, X. Zhang, Z. Dai,
X. Wang, X. Xiao, C. Jiang, Significant Radiation Tolerance and Moderate
Reduction in Thermal Transport of a Tungsten Nanofilm by Inserting
Monolayer Graphene, Adv Mater 29(3) (2017).

{[}27{]} O. El-Atwani, N. Li,
M. Li, A. Devaraj, J.K.S. Baldwin, M.M. Schneider, D. Sobieraj, J.S.
Wróbel, D. Nguyen-Manh, S.A. Maloy, E. Martinez, Outstanding radiation
resistance of tungsten-based high-entropy alloys, Science Advances 5(3)
(2019) eaav2002.

{[}28{]} Y.C. Wu, Q.Q. Hou, L.M. Luo, X. Zan, X.Y. Zhu,
P. Li, Q. Xu, J.G. Cheng, G.N. Luo, J.L. Chen, Preparation of
ultrafine-grained/nanostructured tungsten materials: An overview,
Journal of Alloys and Compounds 779 (2019) 926-941.

{[}29{]} Y. Xu, C.
Xie, S. Qin, J. Song, Q. Li, S. Zhao, G. Liu, T. Wang, Y. Yu, G. Luo,
Preliminary R\&D on flat-type W/Cu plasma-facing materials and
components for Experimental Advanced Superconducting Tokamak, Physica
Scripta 2014(T159) (2014) 014008.

{[}30{]} N. Sun, S. Lang, Y. Zhang, Y.
Xu, H. Liu, G. Li, Properties of electrodeposited tungsten coatings on
graphite substrates for plasma facing components, Journal of Fusion
Energy 35(4) (2016) 660-665.

{[}31{]} R. Potyrailo, K. Rajan, K. Stoewe,
I. Takeuchi, B. Chisholm, H. Lam, Combinatorial and high-throughput
screening of materials libraries: review of state of the art, ACS Comb
Sci 13(6) (2011) 579-633.

{[}32{]} A. Ludwig, Discovery of new materials
using combinatorial synthesis and high-throughput characterization of
thin-film materials libraries combined with computational methods, npj
Computational Materials 5(1) (2019).

{[}33{]} D. You, H. Zhang, S.
Ganorkar, T. Kim, J. Schroers, J.J. Vlassak, D. Lee, Electrical
resistivity as a descriptor for classification of amorphous versus
crystalline phases of alloys, Acta Materialia 231 (2022) 117861.

{[}34{]}
K. Kim, S. Park, T. Kim, Y. Park, G.-D. Sim, D. Lee, Mechanical,
Electrical Properties and Microstructures of Combinatorial Ni-Mo-W alloy
films, Journal of Alloys and Compounds (2022) 165808.

{[}35{]} W. Jiang,
Y. Zhu, L. Zhang, D.J. Edwards, N.R. Overman, G. Nandipati, W. Setyawan,
C.H. Henager Jr, R.J. Kurtz, Dose rate effects on damage accumulation
and void growth in self-ion irradiated tungsten, Journal of Nuclear
Materials 550 (2021) 152905.

{[}36{]} M. Seo, K. Wang, J.R. Echols, A.L.
Winfrey, Microstructure deformation and near-pore environment of
resolidified tungsten in high heat flux conditions, Journal of Nuclear
Materials 565 (2022) 153725.

{[}37{]} S. Wang, W. Guo, Y. Yuan, N. Gao,
X. Zhu, L. Cheng, X. Cao, E. Fu, L. Shi, F. Gao, Evolution of vacancy
defects in heavy ion irradiated tungsten exposed to helium plasma,
Journal of Nuclear Materials 532 (2020) 152051.

{[}38{]} D.S. Perloff,
Four-point sheet resistance correction factors for thin rectangular
samples, Solid-State Electronics 20(8) (1977) 681-687.

{[}39{]} I.
Miccoli, F. Edler, H. Pfnur, C. Tegenkamp, The 100th anniversary of the
four-point probe technique: the role of probe geometries in isotropic
and anisotropic systems, J Phys Condens Matter 27(22) (2015)
223201.

{[}40{]} R. Franz, G. Wiedemann, Ueber die
Wärme-Leitungsfähigkeit der Metalle, Annalen der Physik 165(8) (1853)
497-531.

{[}41{]} B. Fu, W. Lai, Y. Yuan, H. Xu, W. Liu, Calculation and
analysis of lattice thermal conductivity in tungsten by molecular
dynamics, Journal of Nuclear Materials 427(1-3) (2012) 268-273.

{[}42{]}
S. Cui, M. Simmonds, W. Qin, F. Ren, G.R. Tynan, R.P. Doerner, R. Chen,
Thermal conductivity reduction of tungsten plasma facing material due to
helium plasma irradiation in PISCES using the improved 3-omega method,
Journal of Nuclear Materials 486 (2017) 267-273.

{[}43{]} W.C. Oliver,
G.M. Pharr, An improved technique for determining hardness and elastic
modulus using load and displacement sensing indentation experiments,
Journal of Materials Research 7(6) (1992) 1564-1583.

{[}44{]} A.
Fischer-Cripps, A review of analysis methods for sub-micron indentation
testing, Vacuum 58(4) (2000) 569-585.

{[}45{]} X. Cai, H. Bangert,
Hardness measurements of thin films-determining the critical ratio of
depth to thickness using FEM, Thin Solid Films 264(1) (1995)
59-71.

{[}46{]} N. Astm, Standard practice for neutron radiation damage
simulation by charged-particle irradiation, Annu. B. ASTM Stand. 12
(1996) E521.

{[}47{]} M.J. Banisalman, S. Park, T. Oda, Evaluation of the
threshold displacement energy in tungsten by molecular dynamics
calculations, Journal of Nuclear Materials 495 (2017) 277-284.

{[}48{]}
J.A. Thornton, High rate thick film growth, Annual review of materials
science 7(1) (1977) 239-260.

{[}49{]} O. El-Atwani, K. Hattar, J.A.
Hinks, G. Greaves, S.S. Harilal, A. Hassanein, Helium bubble formation
in ultrafine and nanocrystalline tungsten under different extreme
conditions, Journal of Nuclear Materials 458 (2015) 216-223.

{[}50{]} G.
Wei, J. Li, Y. Li, H. Deng, C. Jiang, F. Ren, A better nanochannel
tungsten film in releasing helium atoms, Journal of Nuclear Materials
532 (2020).

{[}51{]} F. Zhu, D. Wang, N. Gao, H. Peng, Z. Xie, Z. Zhang,
Microstructure evolution and Young's modulus of He-implanted
nanocrystalline tungsten film, Journal of Nuclear Materials 518 (2019)
226-233.

{[}52{]} Z. Shang, J. Ding, C. Fan, D. Chen, J. Li, Y. Zhang, Y.
Wang, H. Wang, X. Zhang, He ion irradiation response of a gradient T91
steel, Acta Materialia 196 (2020) 175-190.

{[}53{]} A. Luo, D.L.
Jacobson, K.S. Shin, Solution softening mechanism of iridium and rhenium
in tungsten at room temperature, International Journal of Refractory
Metals and Hard Materials 10(2) (1991) 107-114.

{[}54{]} J.R. Stephens,
W.R. Witzke, Alloy softening in group via metals alloyed with rhenium,
Journal of the Less Common Metals 23(4) (1971) 325-342.

{[}55{]} J.
Stephens, Dislocation structure in single-crystal tungsten and tungsten
alloys, Metallurgical and Materials Transactions B 10(1) (1970)
1293.

{[}56{]} E. Tejado, P.A. Carvalho, A. Munoz, M. Dias, J.B. Correia,
U.V. Mardolcar, J.Y. Pastor, The effects of tantalum addition on the
microtexture and mechanical behaviour of tungsten for ITER applications,
Journal of Nuclear Materials 467 (2015) 949-955.

{[}57{]} Z. Wang, Y.
Yuan, K. Arshad, J. Wang, Z. Zhou, J. Tang, G.-H. Lu, Effects of
tantalum concentration on the microstructures and mechanical properties
of tungsten-tantalum alloys, Fusion Engineering and Design 125 (2017)
496-502.

{[}58{]} J. Li, B. Lu, Y. Zhang, H. Zhou, G. Hu, R. Xia,
Nanoindentation response of nanocrystalline copper via molecular
dynamics: Grain-size effect, Materials Chemistry and Physics 241
(2020).

{[}59{]} F. Hofmann, D.R. Mason, J.K. Eliason, A.A. Maznev, K.A.
Nelson, S.L. Dudarev, Non-Contact Measurement of Thermal Diffusivity in
Ion-Implanted Nuclear Materials, Sci Rep 5 (2015) 16042.

{[}60{]} R.
Taylor, R. Finch, The specific heats and resistivities of molybdenum,
tantalum, and rhenium, Journal of the Less Common Metals 6(4) (1964)
283-294.

{[}61{]} M. Zhao, W. Pan, C. Wan, Z. Qu, Z. Li, J. Yang, Defect
engineering in development of low thermal conductivity materials: A
review, Journal of the European Ceramic Society 37(1) (2017)
1-13.

{[}62{]} C. Kittel, Introduction to Solid State Physics, 6th
edition, Wiley, New York1986.

{[}63{]} A. Debelle, A. Michel, G. Abadias,
C. Jaouen, Ion-irradiation induced stress relaxation in metallic thin
films and multilayers grown by ion beam sputtering, Nuclear Instruments
and Methods in Physics Research Section B: Beam Interactions with
Materials and Atoms 242(1-2) (2006) 461-465.

{[}64{]} A. Debelle, G.
Abadias, A. Michel, C. Jaouen, V. Pelosin, Growth stress buildup in ion
beam sputtered Mo thin films and comparative study of stress relaxation
upon thermal annealing or ion irradiation, Journal of Vacuum Science \&
Technology A: Vacuum, Surfaces, and Films 25(5) (2007).

{[}65{]} K.
Hlushko, A. Mackova, J. Zalesak, M. Burghammer, A. Davydok, C. Krywka,
R. Daniel, J. Keckes, J. Todt, Ion irradiation-induced localized stress
relaxation in W thin film revealed by cross-sectional X-ray
nanodiffraction, Thin Solid Films 722 (2021).

{[}66{]} R.-Y. Zheng, W.-R.
Jian, I.J. Beyerlein, W.-Z. Han, Atomic-scale hidden point-defect
complexes induce ultrahigh-irradiation hardening in tungsten, Nano
Letters 21(13) (2021) 5798-5804.

{[}67{]} C.S. Becquart, C. Domain,
Migration energy of He in W revisited by ab initio calculations, Phys
Rev Lett 97(19) (2006) 196402.

{[}68{]} W. Wilson, Theory of small
clusters of helium in metals, Radiation Effects 78(1-4) (1983)
11-24.

{[}69{]} Y.-W. You, D. Li, X.-S. Kong, X. Wu, C. Liu, Q. Fang, B.
Pan, J. Chen, G.-N. Luo, Clustering of H and He, and their effects on
vacancy evolution in tungsten in a fusion environment, Nuclear Fusion
54(10) (2014) 103007.

{[}70{]} Y.-W. You, J. Sun, X.-S. Kong, X. Wu, Y.
Xu, X. Wang, Q. Fang, C. Liu, Effects of self-interstitial atom on
behaviors of hydrogen and helium in tungsten, Physica Scripta 95(7)
(2020) 075708.

{[}71{]} Y.-W. You, J. Sun, X.-S. Kong, X. Wu, Y. Xu, X.P.
Wang, Q.F. Fang, C.S. Liu, Effects of self-interstitial atom on
behaviors of hydrogen and helium in tungsten, Physica Scripta 95(7)
(2020).

{[}72{]} T. Ungár, Microstructural parameters from X-ray
diffraction peak broadening, Scripta Materialia 51(8) (2004)
777-781.

{[}73{]} Z. Guo, L. Wang, X.-Z. Wang, Additive manufacturing of
W-12Ta(wt\%) alloy: Processing and resulting mechanical properties,
Journal of Alloys and Compounds 868 (2021) 159193.

{[}74{]} G. Kim, X.
Chai, L. Yu, X. Cheng, D.S. Gianola, Interplay between grain boundary
segregation and electrical resistivity in dilute nanocrystalline Cu
alloys, Scripta Materialia 123 (2016) 113-117.

{[}75{]} D. Chen, Y. Tong,
H. Li, J. Wang, Y.L. Zhao, A. Hu, J.J. Kai, Helium accumulation and
bubble formation in FeCoNiCr alloy under high fluence He+ implantation,
Journal of Nuclear Materials 501 (2018) 208-216.

{[}76{]} F.-Y. Yue,
Y.-H. Li, Q.-Y. Ren, F.-F. Ma, Z.-Z. Li, H.-B. Zhou, H. Deng, Y. Zhang,
G.-H. Lu, Suppressing/enhancing effect of rhenium on helium clusters
evolution in tungsten: Dependence on rhenium distribution, Journal of
Nuclear Materials 543 (2021).

{[}77{]} Y. Chen, J. Fang, X. Liao, N. Gao,
W. Hu, H.-B. Zhou, H. Deng, Energetics and diffusional properties of
helium in W-Ta systems studied by a new ternary potential, Journal of
Nuclear Materials 549 (2021).

{[}78{]} T. Hwang, A. Hasegawa, K. Tomura,
N. Ebisawa, T. Toyama, Y. Nagai, M. Fukuda, T. Miyazawa, T. Tanaka, S.
Nogami, Effect of neutron irradiation on rhenium cluster formation in
tungsten and tungsten-rhenium alloys, Journal of Nuclear Materials 507
(2018) 78-86.

{[}79{]} J. Fu, Y. Chen, J. Fang, N. Gao, W. Hu, C. Jiang,
H.-B. Zhou, G.-H. Lu, F. Gao, H. Deng, Molecular dynamics simulations of
high-energy radiation damage in W and W--Re alloys, Journal of Nuclear
Materials 524 (2019) 9-20.

{[}80{]} C.-H. Huang, L. Gharaee, Y. Zhao, P.
Erhart, J. Marian, Mechanism of nucleation and incipient growth of Re
clusters in irradiated W-Re alloys from kinetic Monte Carlo simulations,
Physical Review B 96(9) (2017).

{[}81{]} A. Xu, D.E.J. Armstrong, C.
Beck, M.P. Moody, G.D.W. Smith, P.A.J. Bagot, S.G. Roberts,
Ion-irradiation induced clustering in W-Re-Ta, W-Re and W-Ta alloys: An
atom probe tomography and nanoindentation study, Acta Materialia 124
(2017) 71-78.

{[}82{]} Y.-W. You, X.-S. Kong, X. Wu, C.S. Liu, Q.F. Fang,
J.L. Chen, G.N. Luo, Clustering of transmutation elements tantalum,
rhenium and osmium in tungsten in a fusion environment, Nuclear Fusion
57(8) (2017).

{[}83{]} M.H. Cui, Z.G. Wang, L.L. Pang, T.L. Shen, C.F.
Yao, B.S. Li, J.Y. Li, X.Z. Cao, P. Zhang, J.R. Sun, Y.B. Zhu, Y.F. Li,
Y.B. Sheng, Temperature dependent defects evolution and hardening of
tungsten induced by 200keV He-ions, Nuclear Instruments and Methods in
Physics Research Section B: Beam Interactions with Materials and Atoms
307 (2013) 507-511.

{[}84{]} W.S. Cunningham, J.M. Gentile, O. El-Atwani,
C.N. Taylor, M. Efe, S.A. Maloy, J.R. Trelewicz, Softening due to Grain
Boundary Cavity Formation and its Competition with Hardening in Helium
Implanted Nanocrystalline Tungsten, Sci Rep 8(1) (2018) 2897.

{[}85{]} T.
Wang, G.K.H. Madsen, A. Hartmaier, Atomistic study of the influence of
lattice defects on the thermal conductivity of silicon, Modelling and
Simulation in Materials Science and Engineering 22(3) (2014).

\end{document}


Combinatorial Discovery of Irradiation Damage Tolerant Nano-structured
W-based alloys

Haechan Jo \textsuperscript{a}, Sanghun Park \textsuperscript{a}, Daegun
You \textsuperscript{a}, Sooran Kim \textsuperscript{b}, Dongwoo Lee
\textsuperscript{a, *}

\textsuperscript{a} School of Mechanical Engineering, Sungkyunkwan
University (SKKU), Suwon 16419, Republic of Korea

\textsuperscript{b} Department of Physics Education, Kyungpook National
University (KNU), Daegu 41944, Republic of Korea

*Corresponding Authors Email: dongwoolee@skku.edu

\textbf{\hfill\break
}

\textbf{1. Physical properties and microstructures of the combinatorial
thin film alloys}

The mechanical, electrical, thermal properties, XRD results, TEM
micrographs that are not shown in the manuscript are presented in Fig.
S1-7.

\includegraphics[width=4.8in]{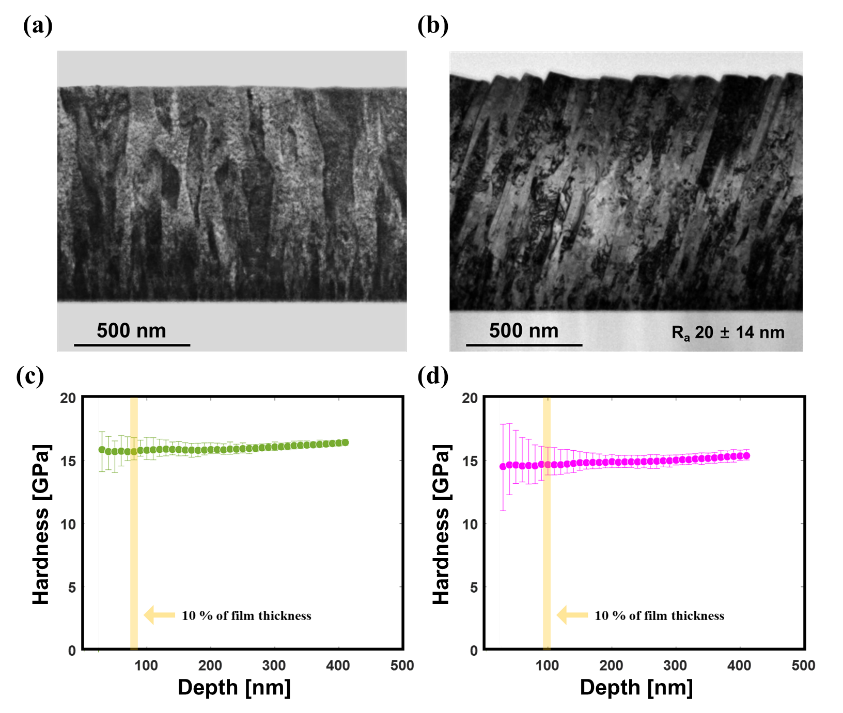}

\textbf{Fig. S1.} Cross-sectional TEM micrographs of pure W films (a)
with a smooth surface and (b) a rough surface (Ra = 20\(\pm\)14 nm). (c,
d) Nano-indentation hardness vs. depth graphs for the samples in (a) and
(b), respectively.

\includegraphics[width=4.8in]{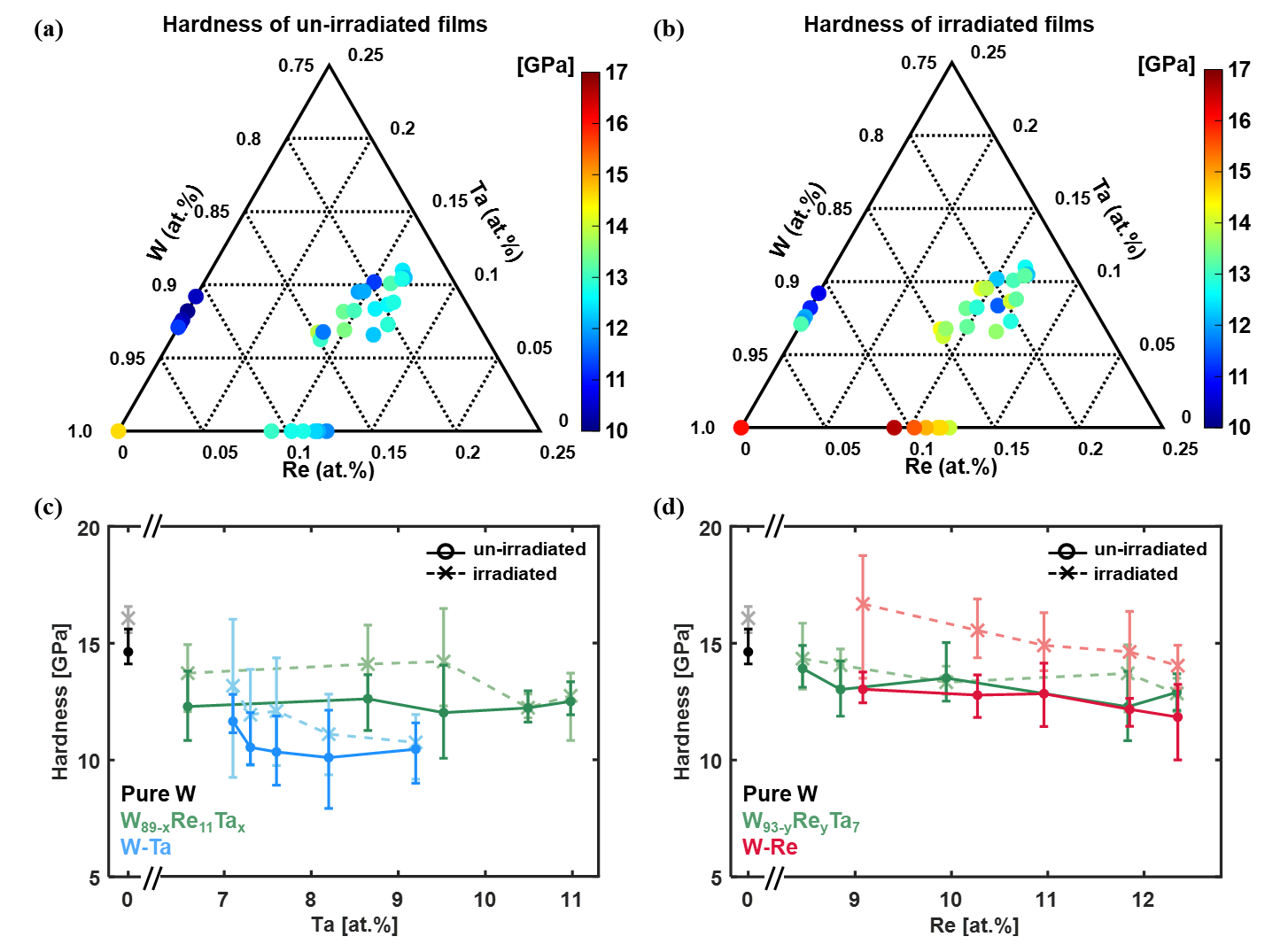}

Fig. S2. The nanoindentation hardness maps of the (a) un-irradiated and
(b) irradiated combinatorial films. (c, d) Hardness of selected
compositions of the W-Ta, W-Re, and W-Re-Ta systems are shown with error
bars.

\includegraphics[width=4.8in]{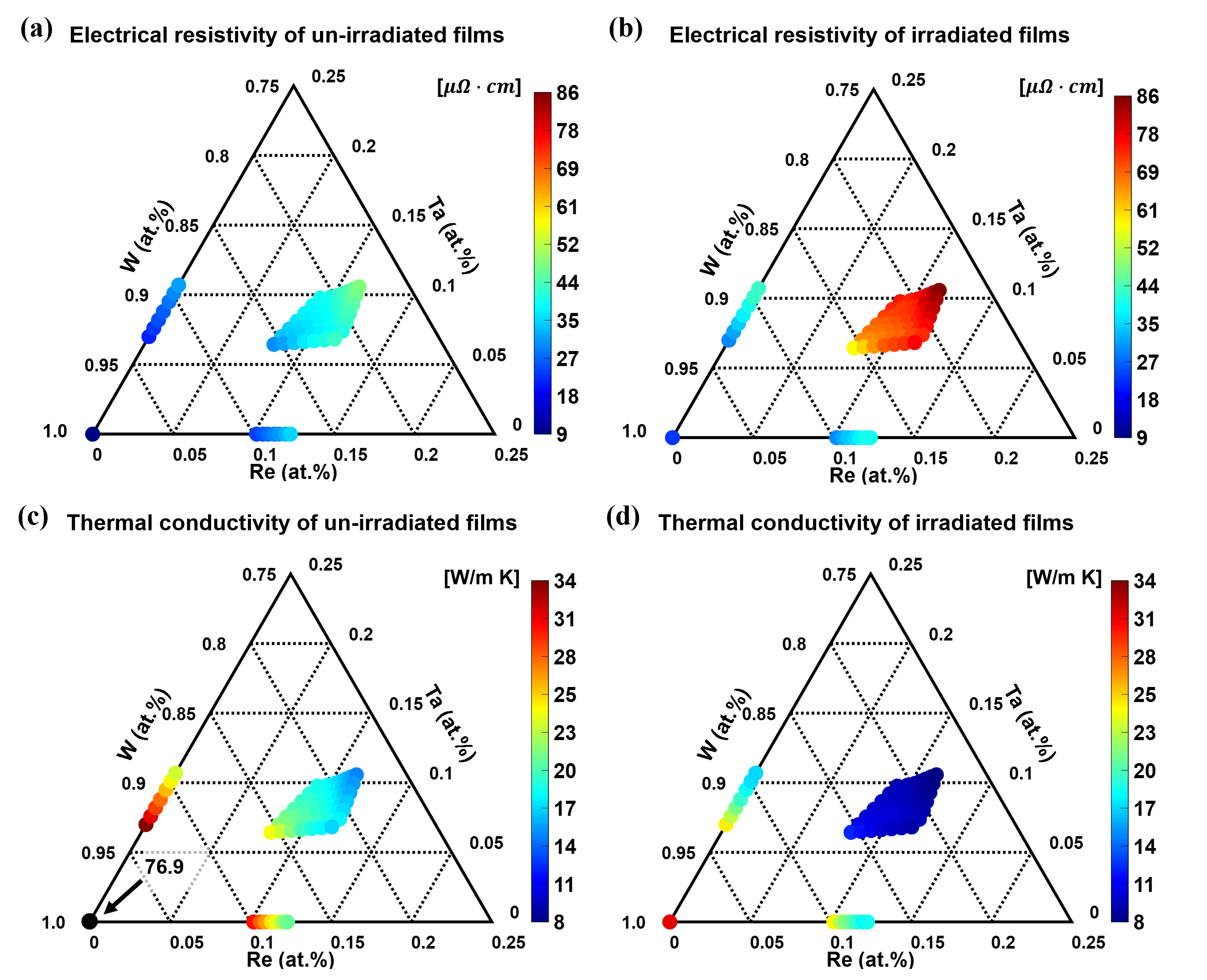}

\textbf{Fig. S3.} Electrical resistivity maps of the (a) un-irradiated
and (b) irradiated combinatorial films. Electronic thermal conductivity
maps of the (c) un-irradiated and (d) irradiated films.

\includegraphics[width=4.8in]{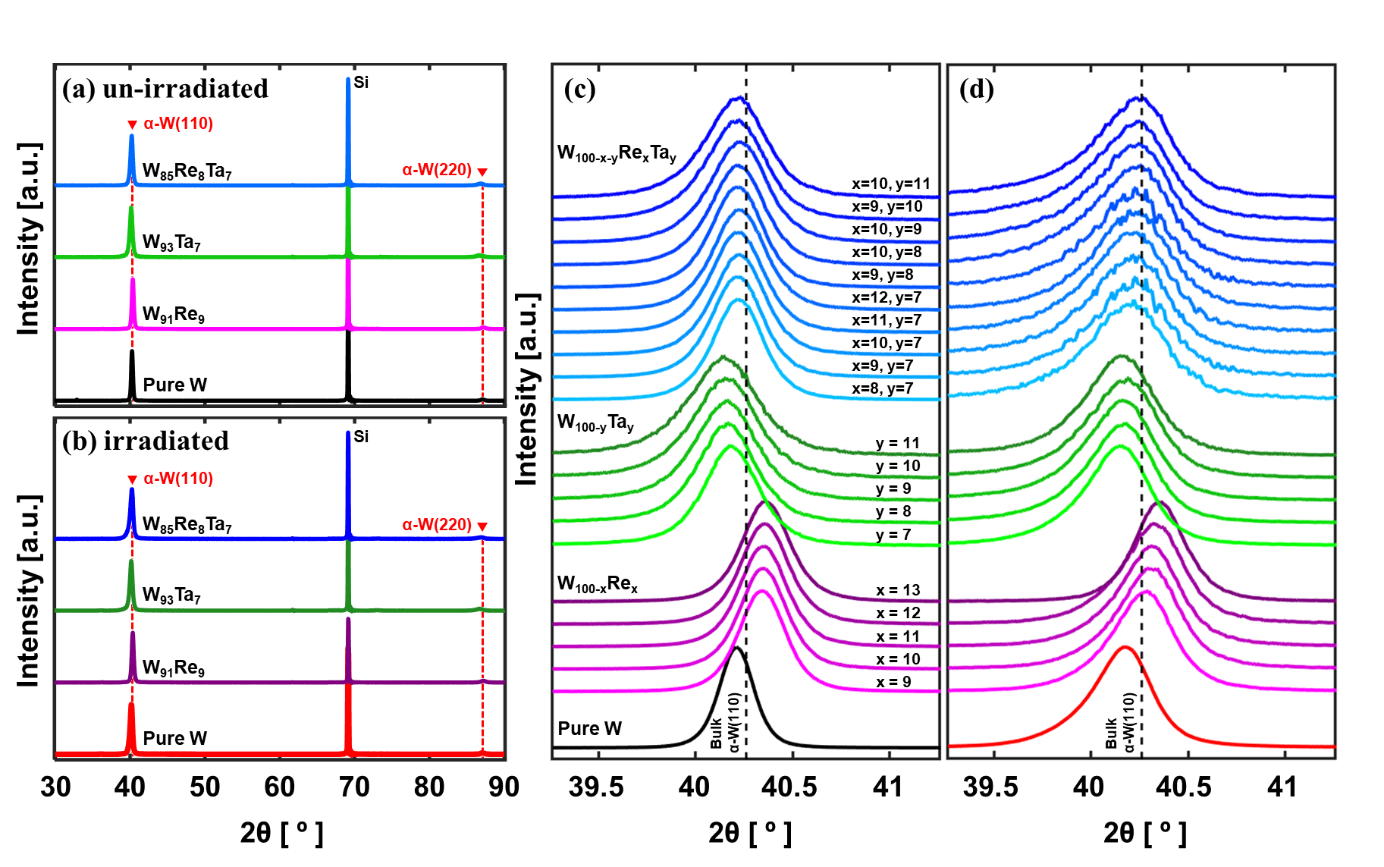}

\textbf{Fig. S4.} XRD spectra of some of the (a) un-irradiated and (b)
irradiated films. A magnified view of the (110) peaks of the (c)
un-irradiated and (d) irradiated films with selected compositions of the
W, W-Re, W-Ta, and W-Re-Ta systems. The lattice parameter of
un-irradiated pure W film was determined as 3.1678 Å (c), which can be
compared with the bulk value of 3.1648 Å. This difference gives 0.095\%
of the out-of-plane residual strain, which corresponds to -0.135 \% of
in-plane residual strain and -0.60 GPa of in-plane residual stress.

\includegraphics[width=4.8in]{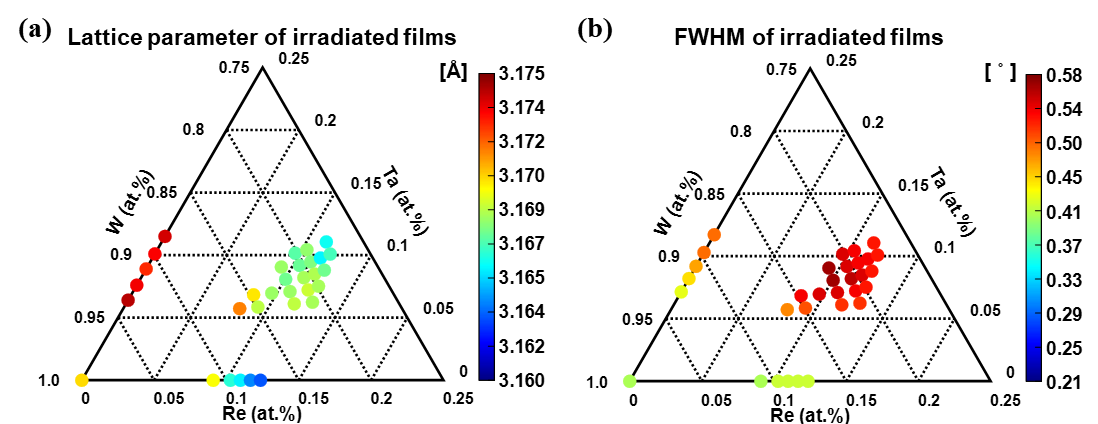}

\textbf{Fig. S5.} (a) The lattice parameter map and (b) the FWHM (full
width at half maximum) map determined from the (110) XRD peaks of the
irradiated films.

\includegraphics[width=4.8in]{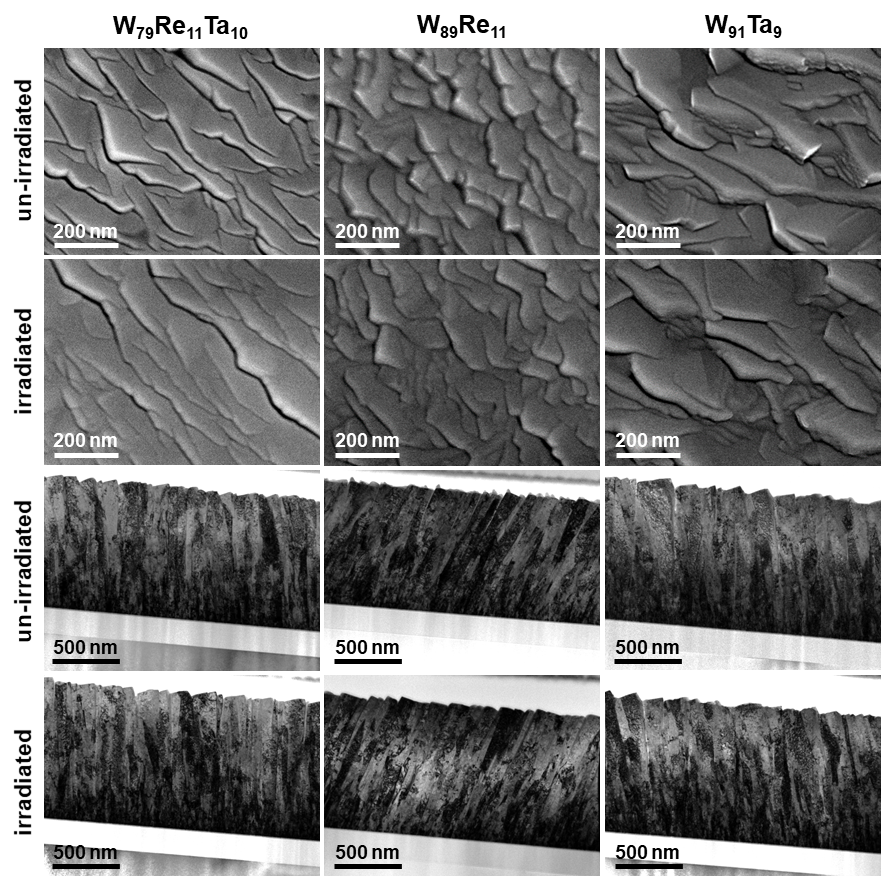}

\begin{quote}
\textbf{Fig. S6.} Surface SEM (top two rows) and bright-field
cross-sectional TEM (bottom two rows) images of the Pure W,
W\textsubscript{89}Re\textsubscript{11},
W\textsubscript{91}Ta\textsubscript{9} and
W\textsubscript{79}Re\textsubscript{11}Ta\textsubscript{10} specimens.
\end{quote}

\includegraphics[width=4.8in]{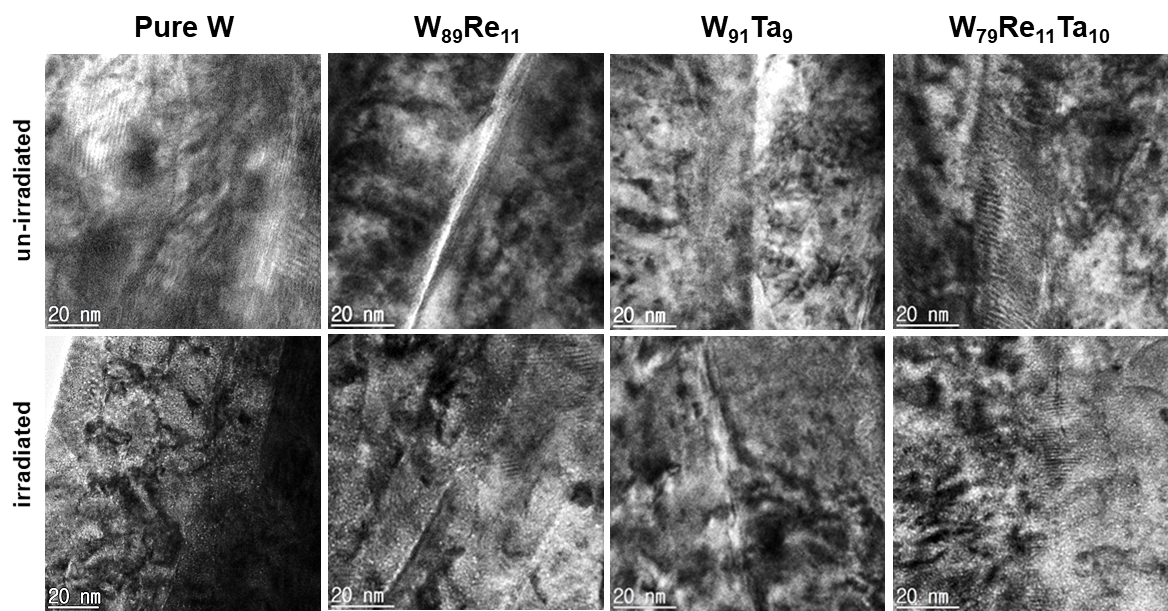}

\begin{quote}
\textbf{Fig. S7.} Cross-sectional TEM micrographs of the Pure W,
W\textsubscript{89}Re\textsubscript{11},
W\textsubscript{91}Ta\textsubscript{9} and
W\textsubscript{79}Re\textsubscript{11}Ta\textsubscript{10} films at
high magnifications.
\end{quote}

\textbf{2. First-principles calculations}

\textbf{2-1. Stabilities of the Ta and Re solutes, and He interstitials}

In this section, we investigate the stabilities of substitutional Ta and
Re atoms, as well as the interstitial He atom by calculating formation
energies (\emph{E\textsubscript{f}}) of each of the atoms. The W
supercell before inserting any point defect is composed of
3\(\times\)3\(\times\)3 BCC unit cells with a lattice constant of 3.185 {\AA} (a total of 54 host atoms). We then inserted the Ta or Re atom to a substitutional position (Fig. S8(a)) and relaxed the structure.
Similarly, a He atom was inserted to the tetrahedral interstitial site
(TIS) or octahedral interstitial site (OIS) (Fig. S8(b) and (c)). The
created structures were relaxed by DFT simulations as described in
\emph{Experimental section}.

\includegraphics[width=4.8in]{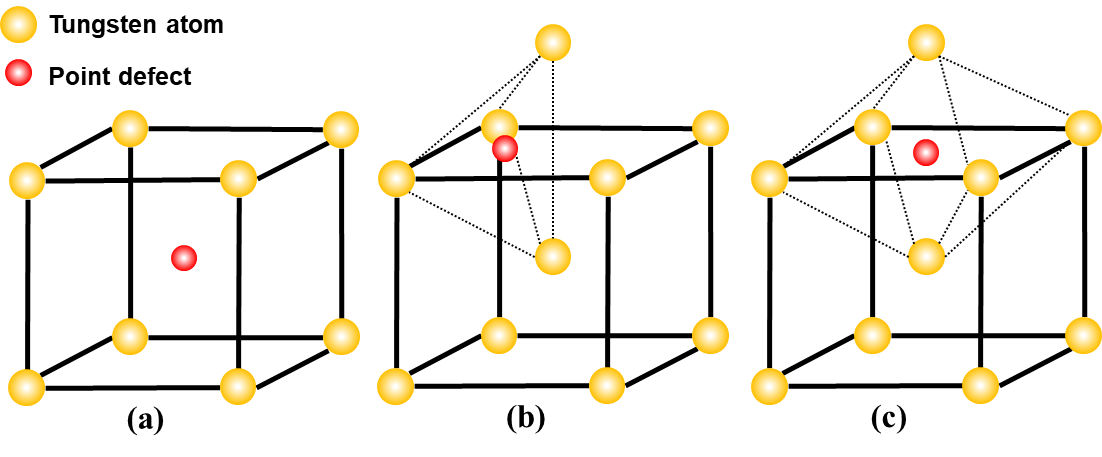}

\textbf{Fig. S8.} Schematic illustrates of the atomic positions of (a)
the substitutional site and the (b) tetrahedral, (c) octahedral
interstitial sites.

The formation energies of each structure were then calculated by using:

\begin{longtable}[]{@{}ll@{}}
\toprule
\endhead
\(E_{f} = E_{\text{tot}}^{X} - E_{\text{tot}}^{p} - \ \sum_{i}^{}{n_{i}\mu_{i}}\ \){[}2{]},
& (Equation S1) \\
\bottomrule
\end{longtable}

where \(E_{\text{tot}}^{X}\) is the total energies of the relaxed
supercell including the point defect, \(E_{\text{tot}}^{p}\) is the
total energies of the crystal without any defect (pure W), \(\mu_{i}\)
is the chemical potential of element species \emph{i}, and \(n_{i}\) is
the number of elements of species \emph{i} added to (\(n_{i} > 0)\) or
removed from (\(n_{i} < 0)\) the supercell. The adopted chemical
potential for W, Ta, Re, and He are -12.96, -11.85, -12.44 eV and
-0.0071 eV, respectively {[}3{]}. The formation energies of
substitutional Ta (\emph{Ta\textsubscript{Sub}}) and Re
(\emph{Re\textsubscript{Sub}}) respectively were calculated as -0.44 and
0.25 eV. The calculated formation energies of the interstitial He at OIS
(\emph{He\textsubscript{OIS}}) and at TIS (\emph{He\textsubscript{TIS}})
are 6.45 and 6.24 eV. These results agree reasonably well with
references (Table S1).

\textbf{Table S1.} Formation energies (eV) of the point defects in W.

\begin{longtable}[]{@{}lllll@{}}
\toprule
& \emph{Ta\textsubscript{Sub}} & \emph{Re\textsubscript{Sub}} &
\emph{He\textsubscript{OIS}} & \emph{He\textsubscript{TIS}} \\
\midrule
\endhead
Present study & - 0.44 & 0.25 & 6.45 & 6.24 \\
Reference {[}2{]} & \emph{-} & \emph{-} & \emph{6.39} & \emph{6.26} \\
Reference {[}4{]} & \emph{-0.45} & \emph{0.18} & \emph{6.18} &
\emph{-} \\
Reference {[}5{]} & \emph{-0.47} & \emph{0.17} & \emph{-} & \emph{-} \\
Reference {[}6{]} & \emph{-} & \emph{-} & \emph{6.48} & \emph{6.23} \\
Reference {[}7{]} & - & - & 6.42 & 6.19 \\
\bottomrule
\end{longtable}

Next, structures with higher solute (Ta or Re) concentrations were
created by inserting six of solutes at random positions of the W
supercell, making the overall solute concentration of 11 \%. For the W,
W\textsubscript{89}Ta\textsubscript{11}, and
W\textsubscript{89}Re\textsubscript{11} supercells, He atoms were
inserted to random TIS as it is more stable than OIS (Table S1). The
number of He atoms inserted varied from one to four, making the overall
He concentration from 1.8 to 6.9 \%. The formation energies of the
interstitial He in the W, W\textsubscript{89}Ta\textsubscript{11}, and
W\textsubscript{89}Re\textsubscript{11} systems can be calculated by:

\begin{longtable}[]{@{}ll@{}}
\toprule
\endhead
\(E_{f} = (E_{\text{tot}}^{{W_{54 - m}\text{Sol}}_{m}\text{He}_{n}} - E_{\text{tot}}^{{W_{54 - m}\text{Sol}}_{m}} - n_{\text{He}} \bullet \mu_{\text{He}})/n_{\text{He}}\ \){[}8{]},
& (Equation S2) \\
\bottomrule
\end{longtable}

where \(E_{\text{tot}}^{{W_{54 - m}\text{Sol}}_{m}\text{He}_{n}}\) and
\(E_{\text{tot}}^{{W_{54 - m}\text{Sol}}_{m}}\) are the total energies
of the supercells with and without TIS He atoms in the
W\textsubscript{54-m}Sol\textsubscript{m} systems and m,
\(n_{\text{He}}\) represent the numbers of the solutes and He atoms (m =
0, for the pure W case), respectively. Calculated \(E_{f}\) as a
function of He concentration can be seen in manuscript (see Fig. 5(a)).

\textbf{2-2. Structural analyses}

For the relaxed structures of W\textsubscript{100-x}He\textsubscript{x},
(W\textsubscript{89}Ta\textsubscript{11})\textsubscript{100-x}He\textsubscript{x},
(W\textsubscript{89}Re\textsubscript{11})\textsubscript{100-x}, radial
distribution functions (RDFs) were determined as shown in Fig. S9. From
the second peak of the RDF, the lattice parameter of the bcc structure
can be determined (Fig. S9(a)). Also, the periodicity of the lattice can
be analyzed by calculating the FWHM (full width at half maximum) of an
RDF peak. Considering the second peak for the FWHM calculations, we find
that the FWHM increases as the concentration of interstitial He
increases (Fig. S9(b) \textasciitilde{} (d)).

\includegraphics[width=4.8in]{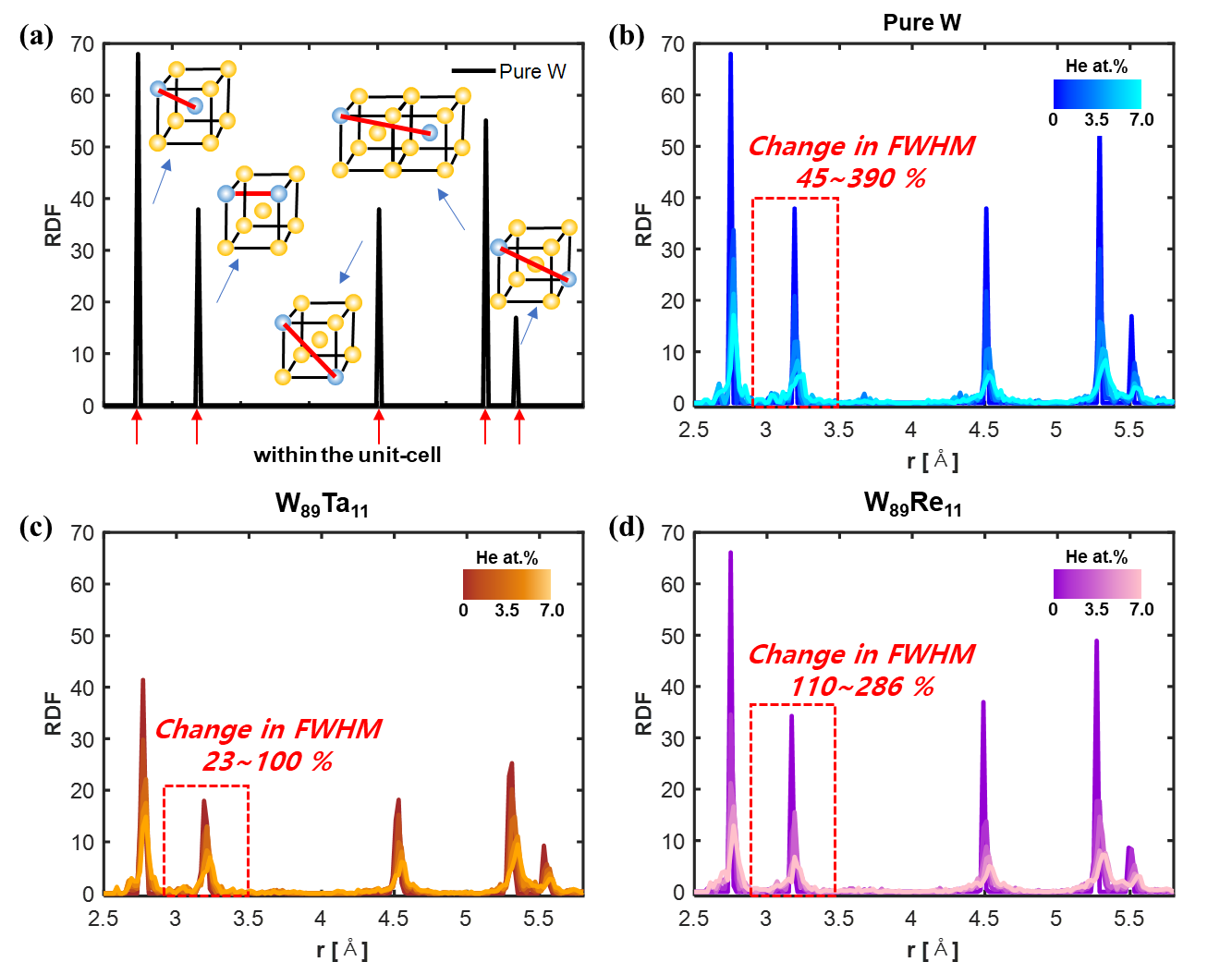}

\textbf{Fig. S9.} RDF curves of (a) Pure W, (b)
W\textsubscript{100-x}He\textsubscript{x}, (c)
(W\textsubscript{89}Ta\textsubscript{11})\textsubscript{100-x}He\textsubscript{x}
and (d) (W\textsubscript{89}Re\textsubscript{11})\textsubscript{100-x}.

\includegraphics[width=4.8in]{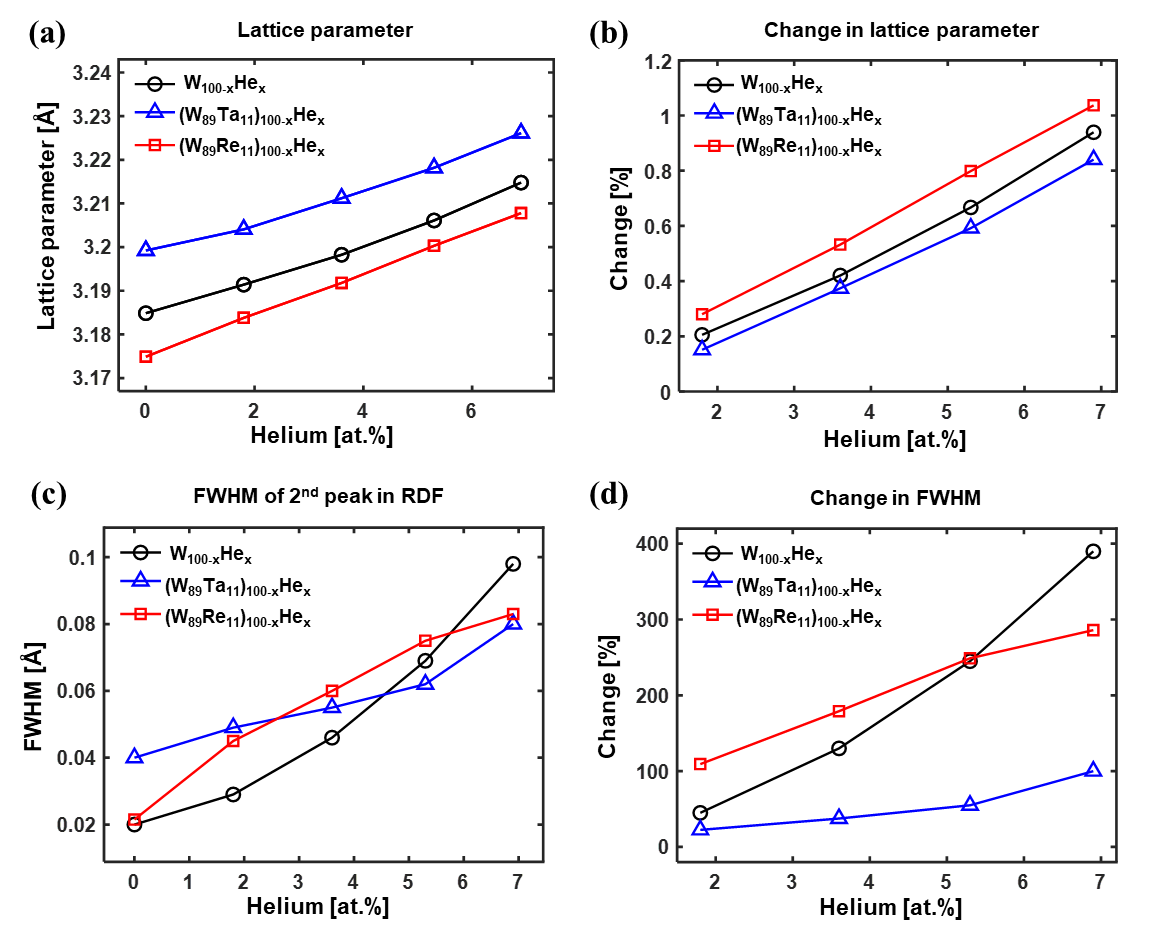}

\textbf{Fig. S10.} (a) The lattice parameter, (b) the change in the
lattice parameter, (c) the FWHM, and (d) the change in FWHM of
W\textsubscript{100-x}He\textsubscript{x},
(W\textsubscript{89}Ta\textsubscript{11})\textsubscript{100-x}He\textsubscript{x},
and (W\textsubscript{89}Re\textsubscript{11})\textsubscript{100-x}
supercells.

The calculated lattice parameters and the FWHM from the RDF curves are
shown in Fig. S9 and S10. Before insertion of any He atoms,
W\textsubscript{89}Ta\textsubscript{11} has the largest lattice
parameter (3.199 Å), followed by pure W (3.185 Å) and
W\textsubscript{89}Re\textsubscript{11} (3.175 Å) (Fig. S10(a)). As the
interstitial He atoms at TIS are inserted, lattice swelling occurs for
all the systems considered (Fig. S10(a)). The change rate in the lattice
parameter due to He insertion is found smallest in the
W\textsubscript{89}Ta\textsubscript{11} system, and is largest in the
W\textsubscript{89}Re\textsubscript{11} systems.

Before He is inserted, the FHWM value is found largest for
W\textsubscript{89}Ta\textsubscript{11}, and the values of W and
W\textsubscript{89}Re\textsubscript{11} are similar (Fig. S10(c)),
indicating that the substitutional Ta atoms distorts the lattice the
most significantly. The FWHM increases with He concentration for all the
systems. The change rate of FWHM due to He concentration is smallest for
the W\textsubscript{89}Ta\textsubscript{11} system (Fig. S10(d)).

\emph{\hfill\break
}

\textbf{References}

{[}1{]} J. Das, A. Chakraborty, T. Bagchi, B. Sarma, Improvement of
machinability of tungsten by copper infiltration technique,
International Journal of Refractory Metals and Hard Materials 26(6)
(2008) 530-539.

{[}2{]} Y.-W. You, J. Sun, X. Wu, Y. Xu, T. Zhang, T.
Hao, Q.F. Fang, C.S. Liu, Interplay of solute-mixed self-interstitial
atoms and substitutional solutes with interstitial and substitutional
helium atoms in tungsten-transition metal alloys, Nuclear Fusion 59(2)
(2019) 026002.

{[}3{]} A. Jain, S.P. Ong, G. Hautier, W. Chen, W.D.
Richards, S. Dacek, S. Cholia, D. Gunter, D. Skinner, G. Ceder,
Commentary: The Materials Project: A materials genome approach to
accelerating materials innovation, APL materials 1(1) (2013)
011002.

{[}4{]} X. Wu, X.-S. Kong, Y.-W. You, C.S. Liu, Q.F. Fang, J.-L.
Chen, G.N. Luo, Z. Wang, Effects of alloying and transmutation
impurities on stability and mobility of helium in tungsten under a
fusion environment, Nuclear Fusion 53(7) (2013) 073049.

{[}5{]} L.
Gharaee, P. Erhart, A first-principles investigation of interstitial
defects in dilute tungsten alloys, Journal of Nuclear Materials 467
(2015) 448-456.

{[}6{]} S.-C. Lee, J.-H. Choi, J.G. Lee, Energetics of He
and H atoms with vacancies in tungsten: First-principles approach,
Journal of Nuclear Materials 383(3) (2009) 244-246.

{[}7{]} X.T. Zu, L.
Yang, F. Gao, S.M. Peng, H.L. Heinisch, X.G. Long, R.J. Kurtz,
Properties of helium defects in bcc and fcc metals investigated with
density functional theory, Physical Review B 80(5) (2009).

{[}8{]} D.
Lee, J.J. Vlassak, K. Zhao, First-Principles Theoretical Studies and
Nanocalorimetry Experiments on Solid-State Alloying of Zr-B, Nano
Letters 15(10) (2015) 6553-8.